\documentclass[longauth]{aa}
\usepackage{graphicx}

\usepackage{txfonts}
\usepackage[breaklinks, colorlinks, citecolor=blue, linkcolor=blue]{hyperref}
\usepackage{natbib}
\bibpunct{(}{)}{;}{a}{}{,} 
\usepackage{xspace}
\usepackage{tabularx}

\newcommand\tess{{\it TESS}\xspace}

\newcommand\gaia{{\it Gaia}\xspace}

\newcommand{\eref}[1]{Equation~(\ref{#1})}
\newcommand{\fref}[1]{Figure~\ref{#1}}
\newcommand{\tref}[1]{Table~\ref{#1}}

\newcommand{\masseighteen}{$2.3 \pm 0.2$~M$\rm_J$\xspace}
\newcommand{\radeighteen}{$1.14 \pm 0.02$~R$\rm_J$\xspace}
\newcommand{\pereighteen}{$4.860674 \pm 0.000005$~d\xspace}
\newcommand{\ecceighteen}{$0.043\pm0.008$\xspace}

\newcommand{\masstwenty}{$4.4 \pm 0.3$~M$\rm_J$\xspace}
\newcommand{\radtwenty}{$1.117 \pm 0.009$~R$\rm_J$\xspace}

\newcommand{\ecctwenty}{$0.41\pm0.02$\xspace}
\newcommand{\lamtwenty}{$9^{+33}_{-31}$~$^\circ$\xspace}
\newcommand{\jitfiestwenty}{$45^{+12}_{-14}$~m~s$^{-1}$\xspace}
\newcommand{\jitfiesptwenty}{$19\pm6$~m~s$^{-1}$\xspace}

\newcommand{\masstwentyone}{$0.82 \pm 0.08$~M$\rm_J$\xspace}
\newcommand{\radtwentyone}{$0.960 \pm 0.012$~R$\rm_J$\xspace}

\usepackage{booktabs}
\usepackage{arydshln}
\usepackage{threeparttable}

\newcommand{\orcid}[1]{\href{https://orcid.org/#1}{\protect\includegraphics[width=8pt]{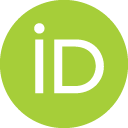}}}

\begin{document} 

\title{Confirmation and characterisation of three giant planets detected by \tess from the FIES/NOT and Tull/McDonald spectrographs}
   \titlerunning{Three giant TOIs}
   \subtitle{}

   \author{
   Emil~Knudstrup\inst{\ref{not},\ref{sac}}\orcid{0000-0001-7880-594X} \and 
   Luisa~M.~Serrano\inst{\ref{torino}}\orcid{0000-0001-9211-3691} \and
   Davide~Gandolfi\inst{\ref{torino}}\orcid{0000-0001-8627-9628} \and
   Simon~H.~Albrecht\inst{\ref{sac}}\orcid{0000-0003-1762-8235} \and
   William~D.~Cochran\inst{\ref{mcd}}\orcid{0000-0001-9662-3496} \and 
   Michael~Endl\inst{\ref{mcd}}\orcid{0000-0002-7714-6310}\and
   Phillip~MacQueen\inst{\ref{mcd}} \and
   Ren\'{e}~Tronsgaard\inst{\ref{dtu}}\orcid{0000-0003-1001-0707} \and 
   Allyson~Bieryla\inst{\ref{harvard}}\orcid{0000-0001-6637-5401} \and 
   Lars~A.~Buchhave\inst{\ref{dtu}}\orcid{0000-0003-1605-5666} \and 
   Keivan~Stassun\inst{\ref{vanderbilt}}\orcid{0000-0002-3481-9052} \and 
   Karen~A.~Collins\inst{\ref{harvard}}\orcid{0000-0001-6588-9574} \and
   Grzegorz~Nowak\inst{\ref{iac},\ref{ull}}\orcid{0000-0002-7031-7754} \and 
   Hans~J.~Deeg\inst{\ref{iac},\ref{ull}}\orcid{0000-0003-0047-4241} \and 
   Khalid~Barkaoui\inst{\ref{liege},\ref{atmos_mit},\ref{iac}}\orcid{0000-0003-1464-9276} \and
   Boris~S.~Safonov\inst{\ref{sternberg}} \and 
   Ivan~A.~Strakhov\inst{\ref{sternberg}} \and 
   Alexandre~A.~Belinski\inst{\ref{sternberg}}\orcid{0000-0003-3469-0989} \and 
   Joseph~D.~Twicken \inst{\ref{seti},\ref{ames}}\orcid{0000-0002-6778-7552} \and 
   Jon~M.~Jenkins \inst{\ref{ames}} \and
   Andrew~W.~Howard\inst{\ref{caltech}}\orcid{0000-0001-8638-0320} \and 
   Howard~Isaacson\inst{\ref{berkely},\ref{toowoomba}}\orcid{0000-0002-0531-1073} \and 
   Joshua~N.~Winn\inst{\ref{princeton}}\orcid{0000-0002-4265-047X} \and 
   Kevin~I.~Collins\inst{\ref{mason}}\orcid{0000-0003-2781-3207}\and
   Dennis~M.~Conti\inst{\ref{var}}\orcid{0000-0003-2239-0567} \and
   Gabor~Furesz\inst{\ref{kavli_mit}} \and
   Tianjun~Gan\inst{\ref{tsing}}\orcid{0000-0002-4503-9705}\and
   John~F.~Kielkopf\inst{\ref{louis}} \orcid{0000-0003-0497-2651}\and
   Bob~Massey\inst{\ref{villa}}\orcid{0000-0001-8879-7138}\and
   Felipe~Murgas\inst{\ref{iac},\ref{ull}} \orcid{0000-0001-9087-1245}\and
   Lauren~G.~Murphy\inst{\ref{kutz}}\orcid{0000-0003-3796-6303}\and
   Enric~Palle\inst{\ref{iac},\ref{ull}} \and
   Samuel~N.~Quinn\inst{\ref{harvard}}\orcid{0000-0002-8964-8377} \and 
   Phillip~A.~Reed \inst{\ref{kutz}}\orcid{0000-0002-5005-1215}\and
   George~R.~Ricker\inst{\ref{kavli_mit}} \orcid{0000-0003-2058-6662}\and
   Sara~Seager\inst{\ref{kavli_mit},\ref{atmos_mit},\ref{aero_mit}} \and
   Bernie~Shiao\inst{\ref{space_tel}} \and
   Richard~P.~Schwarz\inst{\ref{pata}} \orcid{0000-0001-8227-1020}\and
   Gregor~Srdoc\inst{\ref{kotizarovci}} \and
   David~Watanabe\inst{\ref{planet_disc}}
          }

   \authorrunning{Knudstrup et~al.}
   \institute{
   Nordic Optical Telescope, Rambla Jos\'{e} Ana Fern\'{a}ndez P\'{e}rez 7, ES-38711 Bre\~{n}a Baja, Spain\label{not} \and
   Stellar Astrophysics Centre, Department of Physics and Astronomy, Aarhus University, Ny Munkegade 120, DK-8000 Aarhus C, Denmark \email{emil@phys.au.dk}\label{sac} \and
   Dipartimento di Fisica, Università degli Studi di Torino, via Pietro Giuria 1, I-10125, Torino, Italy \label{torino}\and
   McDonald Observatory and Center for Planetary Systems Habitability, The University of Texas, Austin Texas USA\label{mcd}\and
   DTU Space, National Space Institute, Technical University of Denmark, Elektrovej 328, DK-2800 Kgs. Lyngby, Denmark\label{dtu}\and
   Center for Astrophysics ${\rm \mid}$ Harvard {\rm \&} Smithsonian, 60 Garden Street, Cambridge, MA 02138, USA\label{harvard}\and
   Department of Physics and Astronomy, Vanderbilt University, Nashville, TN 37235, USA\label{vanderbilt}\and
   Instituto de Astrof\'isica de Canarias (IAC), E-38205 La Laguna, Tenerife, Spain\label{iac}\and
   Dept. Astrof\'isica, Universidad de La Laguna (ULL), E-38206 La Laguna, Tenerife, Spain\label{ull}\and
   Astrobiology Research Unit, Universit\'e de Li\`ege, All\'ee du 6 Ao\^ut 19C, B-4000 Li\`ege, Belgium\label{liege}\and
   Department of Earth, Atmospheric and Planetary Science, Massachusetts Institute of Technology, 77 Massachusetts Avenue, Cambridge, MA 02139, USA\label{atmos_mit}\and
   Sternberg Astronomical Institute, Lomonosov Moscow State University, 13 Universitetski prospekt, 119992 Moscow, Russia\label{sternberg}\and
   SETI Institute, Mountain View, CA  94043, USA\label{seti}\and
   NASA Ames Research Center, Moffett Field, CA  94035, USA\label{ames}\and
   Department of Astronomy, California Institute of Technology, Pasadena, CA 91125, USA\label{caltech}
   Department of Astronomy, University of California Berkeley, Berkeley CA 94720, USA\label{berkely}\and
   Centre for Astrophysics, University of Southern Queensland, Toowoomba, QLD, Australia\label{toowoomba}\and
   Department of Astrophysical Sciences, Princeton University, Princeton, NJ 08544, USA\label{princeton}\and
   George Mason University, 4400 University Drive, Fairfax, VA, 22030 USA\label{mason}\and
   American Association of Variable Star Observers, 49 Bay State Road, Cambridge, MA 02138, USA\label{var}\and
   Department of Physics and Kavli Institute for Astrophysics and Space Research, Massachusetts Institute of Technology, Cambridge, MA 02139, USA\label{kavli_mit}\and
   Department of Astronomy and Tsinghua Centre for Astrophysics, Tsinghua University, Beijing 100084, China\label{tsing}\and
   Department of Physics and Astronomy, University of Louisville, Louisville, KY 40292, USA\label{louis}\and
   Villa '39 Observatory, Landers, CA 92285, USA\label{villa}\and
   Department of Physical Sciences, Kutztown University, Kutztown, PA 19530, USA\label{kutz}\and
   Department of Aeronautics and Astronautics, MIT, 77 Massachusetts Ave., Cambridge, MA 02139, USA\label{aero_mit}\and
   Space Telescope Science Institute, 3700 San Martin Drive, Baltimore, MD, 21218, USA\label{space_tel}\and
   Patashnick Voorheesville Observatory, Voorheesville, NY 12186, USA\label{pata}\and
   Kotizarovci Observatory, Sarsoni 90, 51216 Viskovo, Croatia\label{kotizarovci}\and
   Planetary Discoveries, Fredericksburg, VA 22405, USA\label{planet_disc}
   }

   \date{Received ...; accepted ...}

  \abstract{
  We report the confirmation and characterisation of TOI-1820~b, TOI-2025~b, and TOI-2158~b, three Jupiter-sized planets on short-period orbits around G-type stars detected by \tess. Through our ground-based efforts using the FIES and Tull spectrographs, we have confirmed these planets and characterised their orbits, and find periods of around $4.9$~d, $8.9$~d, and $8.6$~d for TOI-1820~b, TOI-2025~b, and TOI-2158~b, respectively. The sizes of the planets range from 0.96 to 1.14 Jupiter radii, and their masses are in the range from 0.8 to 4.4 Jupiter masses. For two of the systems, namely TOI-2025 and TOI-2158, we see a long-term trend in the radial velocities, indicating the presence of an outer companion in each of the two systems. For TOI-2025 we furthermore find the star to be well-aligned with the orbit, with a projected obliquity of \lamtwenty. As these planets are all found in relatively bright systems (V$\sim$10.9-11.6 mag), they are well-suited for further studies, which could help shed light on the formation and migration of hot and warm Jupiters.     }

   \keywords{planets and satellites: detection -- techniques: radial velocities -- techniques: photometric --
                 planets and satellites: gaseous planets --
                 planet-star interactions
                 }

   \maketitle
   
%

\section{Introduction}
Giant planets on short-period orbits (also called hot Jupiters) were the first planets to be discovered, and their numbers increased quickly during the first years of exoplanetary science. Their existence itself immediately posed a challenge to planet formation theories, which at the time only had one example, the Solar System. Despite almost three decades of discoveries of hot Jupiters, there is still no consensus on their exact origin channel \citep{Dawson2018}. While it is still unclear whether hot Jupiters can form in situ or not \citep{Batygin2016}, ex situ formation processes require a mechanism responsible for transporting these giant planets from larger separations to the current close-in orbits. 

The two leading hypotheses for such large-scale migration that have been put forward are disc migration and high-eccentricity tidal migration. In the former scenario, the planets exchange angular momentum with the gas and dust particles in the circumstellar disc. As a result, the semi-major axis slowly shrinks, while the orbit remains circular \citep[e.g.][]{Lin1996,Baruteau2014}. In contrast, the latter scenario could result in very eccentric and misaligned orbits, since it involves gravitational interactions with other bodies in the system \citep[e.g.][]{Nagasawa2008,Chatterjee2008}.

The advent of space-based transit search missions has led to the discovery of thousands of new exoplanet candidates \citep[see, e.g.][]{Borucki2010,Huang2013,Livingston2018,Kruse2019}. Combining these discoveries with ground-based spectroscopic follow-up observations leads to a large sample of well-characterised exoplanet systems, including the bulk density of the transiting planets, host star properties, orbital eccentricities, stellar obliquities, and companionship of outer planets or stars \citep[see, e.g.][]{Gandolfi2019,VanEylen2019, Carleo2020,Albrecht2021,Knudstrup2022,Smith2022}.  

Here we report on the discovery of three transiting hot Jupiters: TOI-1820b, TOI-2025b, and TOI-2158b. The transit-like features associated with these systems were detected by the Transiting Exoplanet Survey Satellite \citep[\tess;][]{Ricker15}. We have confirmed these as bona fide planets, and we have characterised the planets and their host systems in terms of masses and orbital eccentricities. For one system (TOI-2025), we additionally performed spectroscopic transit observations and used them to determine the sky-projected spin-orbit obliquity. During the preparation of this manuscript, we became aware of the efforts of another team to announce the discovery of TOI-2025~b \citep{Rodriguez2022}. The results were determined independently, and the communication between the teams were strictly related to the coordination of the manuscripts.

In Section~\ref{sec:tessphot} we describe the \tess photometry and data extraction. We present our ground-based observations, which include both additional photometry and spectroscopic follow-up, as well speckle interferometry, in Section~\ref{sec:data}. In Section~\ref{sec:stelpars} we explain how we obtained stellar parameters for the three systems. The methodology behind our analysis is described in Section~\ref{sec:mcmc}. We discuss our results in Section~\ref{sec:results}, before placing these planets in the context of the population from the literature and drawing our conclusions in Section~\ref{sec:conclusions}.

\begin{figure*}
    \centering
    \includegraphics[width=\textwidth]{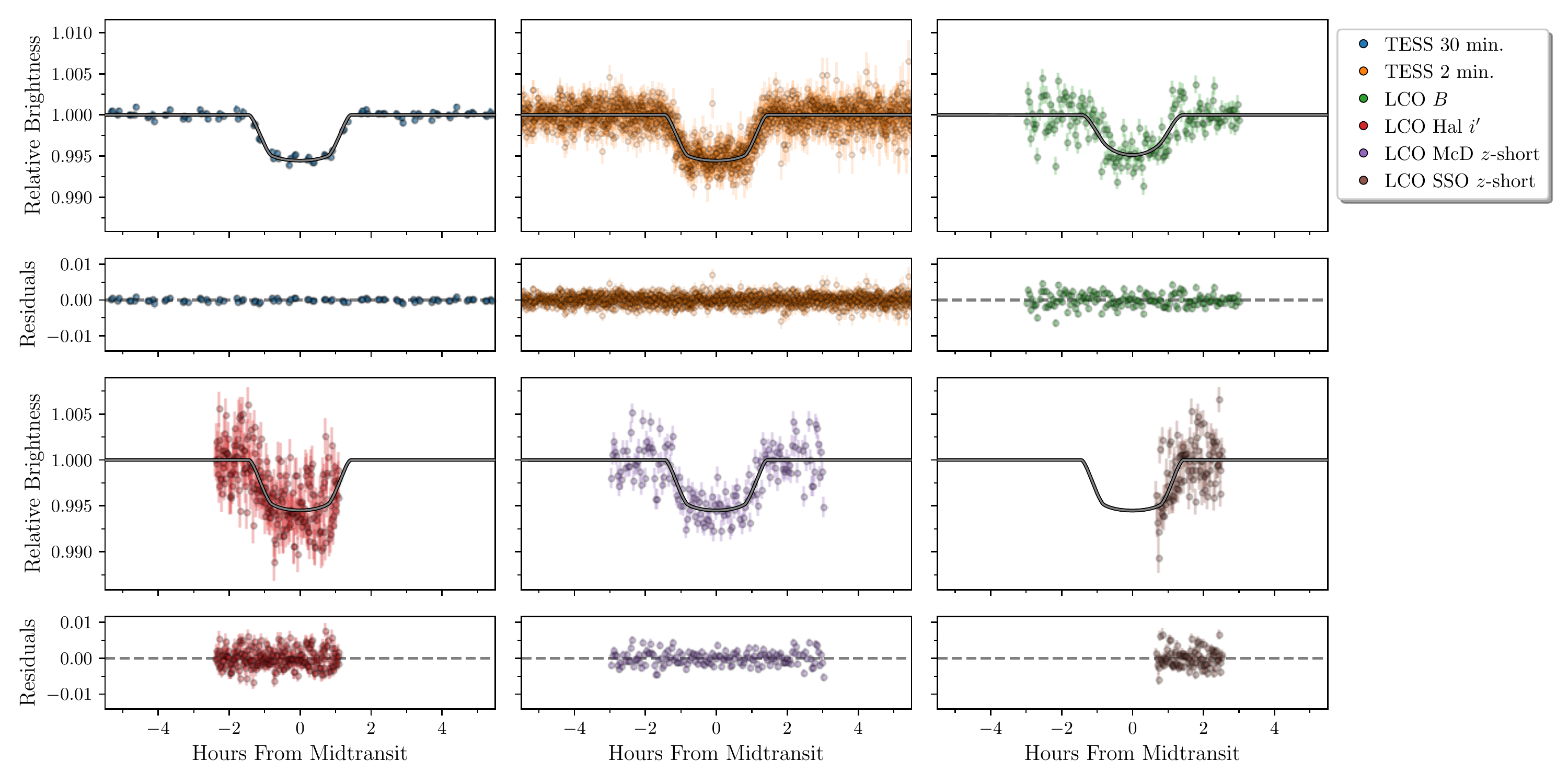}
    \caption{{\bf Photometry for TOI-1820.} Our different photometric observations of TOI-1820 with the best-fitting transit model are shown with a grey line, and the residuals, following the subtraction of the best-fitting model, are shown below.}
    \label{fig:lc_toi1820}
\end{figure*}

\section{TESS photometry of candidate systems}
\label{sec:tessphot}

The transiting planet candidates TOI-1820, TOI-2025, and TOI-2158 were identified by the Massachusetts Institute of Technology (MIT) Quick Look Pipeline \citep[QLP;][]{Huang2020} in a search of light curves extracted from the 30-minute cadence Full Frame Images (FFIs) using the box-least-squares \citep[BLS;][]{Kovacs2002,Hartman2016} algorithm. Transit signals were detected for all three systems, which were then identified as TESS Objects of Interest (TOIs) by the TESS Science Office at MIT \citep{Guerrero21}.

All three targets were subsequently put on the target list for 2-minute cadence. The 2-minute cadence data are processed by the Science Processing Operation Center \citep[SPOC;][]{SPOC} team at the NASA Ames Research Center, where light curves are extracted through simple aperture photometry \citep[SAP;][]{Twicken2010,Morris2020} and processed using the Presearch Data Conditioning \citep[PDC;][]{Smith2012,Stumpe2012,Stumpe2014} algorithm.

We downloaded and extracted all the \tess light curves from the target pixel files using the \texttt{lightkurve} \citep{lightkurve} package, where we use the implemented \texttt{RegressionCorrector} to correct for any background noise. We excluded cadences with severe quality issues\footnote{'default' in \texttt{lightkurve.SearchResult.download}}. We also removed outliers. First we removed the transits from the light curve through a light-curve model using parameters from an initial fit. Next we applied a Savitsky-Golay filter and identified outliers through $5\sigma$ sigma clipping, which we then excluded from the unfiltered light curve with transits. For all three systems, we confirmed the presence of the transit-like features identified by QLP, by performing an independent search using the BLS and the Transit Least Squares \citep[TLS;][]{Hippke19b} algorithm. We furthermore searched for additional transits, without finding hints of any.

\subsection{TOI-1820}
TOI-1820 was observed in Sector 22 (February 18, 2020 and March 18, 2020), with {\it TESS'} camera 1 with a cadence of 30~minutes. TOI-1820 was identified on April 17, 2020 with a signal-to-noise ratio (S/N) of 53. TOI-1820 was observed again in Sector 49 (February 26, 2022 and March 26, 2022) with camera 1, this time with a cadence of 2~minutes. In the top left of \fref{fig:lc_toi1820}, we show the \tess light curve phase folded to the periodic transit signal occurring every 4.860674~d with a depth of $\sim$0.6\%.

\subsection{TOI-2025}
\begin{figure*}
    \centering
    \includegraphics[width=\textwidth]{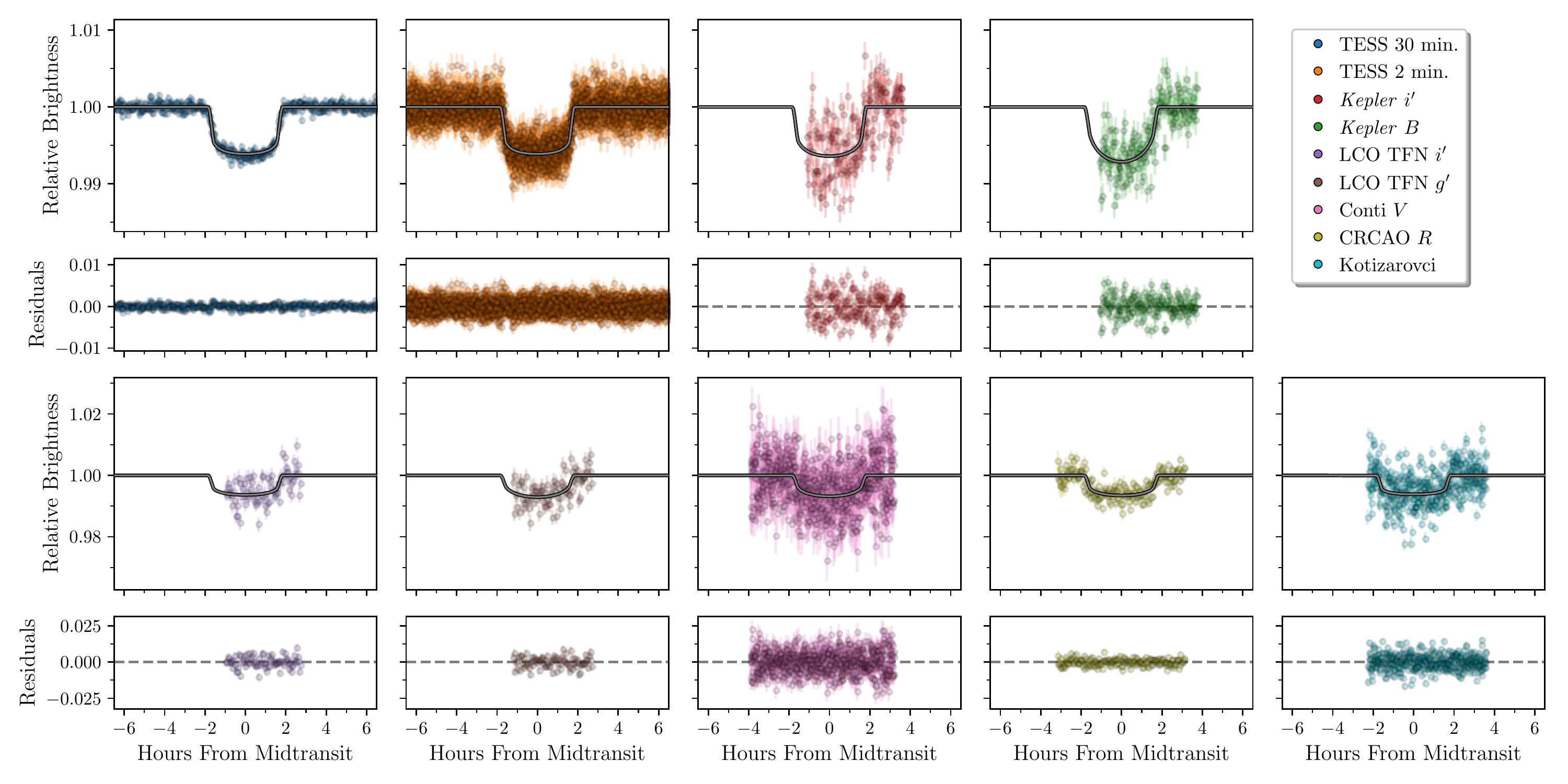}
    \caption{{\bf Photometry for TOI-2025.} Our different photometric observations of TOI-2025 with the best-fitting transit model are shown with a grey line, and the residuals, following the subtraction of the best-fitting model, are shown below.}
    \label{fig:lc_toi2025}
\end{figure*}

TOI-2025 was observed with a 30-minute cadence using {\it TESS'} camera 3 in Sector 14 (July 18, 2019 to August 15, 2019), Sectors 18-20 (November 2, 2019 to January 21, 2020), Sectors 24-26 (April 16, 2020 to July 4, 2020), as well as in 2-minute cadence in Sector 40 (June 24, 2021 to July 23, 2021) and Sector 47 (December 30, 2021 to January 28, 2022), also with camera 3. Since the \tess light curves of TOI-2025 display a periodic 8.872078~d dip of $\sim$0.7\% with a S/N of 151, the candidate was announced as a TOI on June 19, 2020. The two panels on the top left of \fref{fig:lc_toi2025} shows the phase-folded \tess light curves.

\subsection{TOI-2158}
\begin{figure*}
    \centering
    \includegraphics[width=\textwidth]{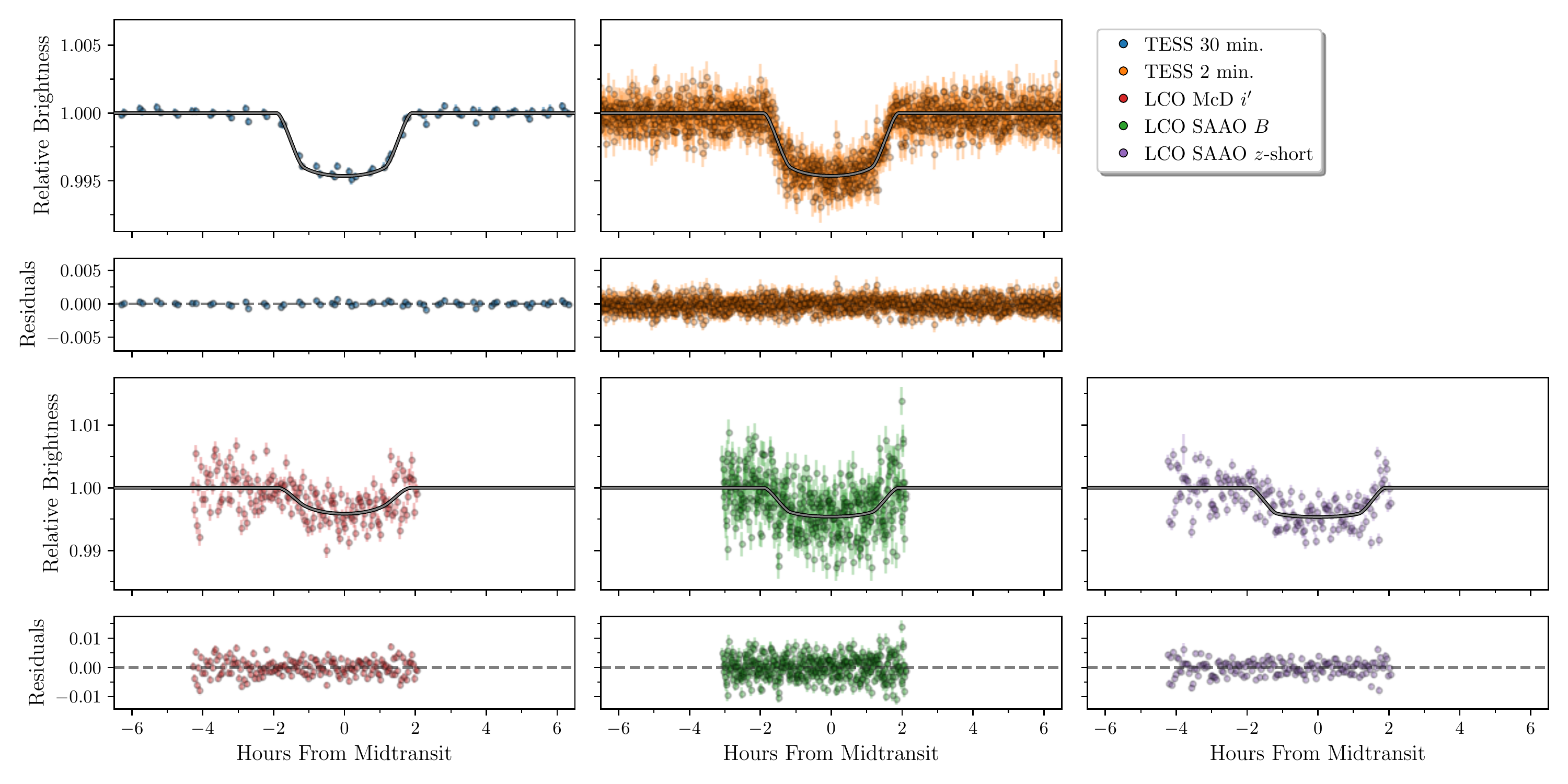}
    \caption{{\bf Photometry for TOI-2158.} Our different photometric observations of TOI-2158 with the best-fitting transit model are shown with a grey line, and the residuals, following the subtraction of the best-fitting model, are shown below.}
    \label{fig:lc_toi2158}
\end{figure*}

TOI-2158 was observed with {\it TESS'} camera 1 during Sector 26 (June 8, 2020 to July 4, 2020) with a cadence of 30 minutes, and in Sector 40 (June 24, 2021 to July 23, 2021) with a 2-minute cadence. On August 10, 2020, TOI-2158 was announced as a TOI with a S/N of 59. The \tess light curve for TOI-2158 can be seen in the top of \fref{fig:lc_toi2158}, phase folded onto the 8.60077~d signal showing the $\sim$0.5\% decrease in flux. A close-up of the \tess light curves for all three systems can be found in \fref{fig:lc_tess2min}.

\begin{table*}[t]
    \centering
    \caption{{\bf Stellar parameters.}}
    \begin{threeparttable}
    \begin{tabular}{l l c c c}
    \toprule
&  \tess Object of Interest &TOI-1820 & TOI-2025 & TOI-2158\\ 
 &  \tess Input Catalogue & TIC 393831507 & TIC 394050135 & TIC 342642208 \\ 
 & TYCHO-2 & TYC 1991-1863-1 & TYC 4595-797-1 & TYC 1577-691-1\\  \midrule 
$V$\tnote{a} & Tycho $V$ magnitude & 10.90 & 11.60 & 10.89 \\ 
$G$\tnote{b} & Gaia $G$ magnitude & 10.97 & 11.36 & 10.67 \\ 
$\alpha_\mathrm{J2000}$\tnote{b} & Right Ascension & 12:30:44.813 & 18:51:10.861 & 18:27:14.413 \\ 
$\delta_\mathrm{J2000}$\tnote{b} & Declination & 27:27:07.206 & 82:14:43.492 & 20:31:36.793 \\ 
$\mu_\alpha$\tnote{b} & Proper motion in R.A. (mas yr$^{-1}$) & 50.54$\pm$0.08 & 2.79$\pm$0.04 & -44.00$\pm$0.04 \\ 
$\mu_\delta$\tnote{b} & Proper motion in Dec. (mas yr$^{-1}$) & -33.93$\pm$0.08 & -4.52$\pm$0.05 & 7.89$\pm$0.07 \\ 
$\varpi$\tnote{b} & Parallax (mas) & 4.00$\pm$0.06 & 2.95$\pm$0.02 & 5.01$\pm$0.04 \\ 
$\pi$\tnote{b} & Distance (pc) & 250$\pm$4 & 339$\pm$2 & 200$\pm$1 \\ 
$T_\mathrm{eff}$\tnote{c} & Effective temperature (K) & 5734$\pm$50 & 5880$\pm$53 & 5673$\pm$50 \\ 
$\log g$\tnote{c} & Surface gravity (dex) & 4.24$\pm$0.05 & 4.17$\pm$0.06 & 4.19$\pm$0.05 \\ 
$\rm [Fe/H]$\tnote{c} & Metallicity (dex) & 0.14$\pm$0.15 & 0.18$\pm$0.08 & 0.47$\pm$0.08 \\ 
$v\sin i_\star$\tnote{c} & Projected rotational velocity (km s$^{-1}$) & 4.5$\pm$0.8 & 6.0$\pm$0.3 & 3.7$\pm$0.5 \\ 
$A_\mathrm{V}$\tnote{d} & Extinction (mag) & 0.04$\pm$0.02 & 0.10$\pm$0.03 & 0.24$\pm$0.02 \\ 
    $F_\mathrm{bol}$\tnote{c} & Bolometric flux (erg s$^{-1}$ cm$^{-2}$) & $(1.017 \pm 0.018)\times10^{-9}$ & $(7.02 \pm 0.16)\times10^{-10}$ & $(1.540 \pm 0.018)\times 10^{-9}$ \\ 
$R_\star$\tnote{d} & Radius (R$_\odot$) & 1.51$\pm$0.06 & 1.56$\pm$0.03 & 1.41$\pm$0.03 \\ 
$M_\star$\tnote{d} & Mass (M$_\odot$) & 1.04$\pm$0.13 & 1.32$\pm$0.14 & 1.12$\pm$0.12 \\ 
$P_\mathrm{rot}/\sin i$\tnote{d} & Rotation period (days) & 25$\pm$6 & 13.2$\pm$0.7 & 19$\pm$3 \\ 
$P_\mathrm{pred}$\tnote{d} & Predicted rotation period (days) & 40$\pm$2 & - & 43$\pm$3 \\ 
$\log$ R$^{\prime}_\mathrm{HK}$\tnote{e} & Activity & -5.37\tnote{e} & - & -5.06$\pm$0.05 \\ 
$\tau$\tnote{d} & Age (Gyr) & 11$\pm$2 & 1.7$\pm$0.2 & 8$\pm$1 \\ 
$\rho$\tnote{d} & Density (g cm$^{-3}$) & 0.43$\pm$0.07 & 0.49$\pm$0.06 & 0.56$\pm$0.07 \\

\bottomrule
    \end{tabular}
\begin{tablenotes}
    \item  Parameters of the stellar hosts in the three systems of this study. 
    \item[a] Tycho-2 \citep{Hoeg2000}.
    \item[b] \gaia EDR3 \citep{GaiaEDR3}.
    \item[c] This work: SPC.
    \item[d] This work: SED.
    \item[e] This work: HIRES spectra.
\end{tablenotes}
\end{threeparttable}
    \label{tab:targets}
\end{table*}

\section{Ground-based observations}
\label{sec:data}

In addition to \tess\ space-based photometry, we gathered ground-based photometry via the Las Cumbres Observatory Global Telescope \citep[LOCGT;][]{Brown:2013}, as well as ground-based spectroscopic measurements from different telescopes. Reconnaissance spectroscopy was acquired with the High Resolution Echelle Spectrometer \citep[HIRES;][]{Vogt1994} located at the Keck Observatory, the Tillinghast Reflector Echelle Spectrograph \citep[TRES;][]{Furesz2008} situated at the Fred L. Whipple Observatory, Mt. Hopkins, AZ, USA, as well as the FIber-fed Echelle Spectrograph \citep[FIES;][]{Frandsen99,Telting14} at the Nordic Optical Telescope \citep[NOT;][]{NOT2010} of the Roque de los Muchachos observatory, La Palma, Spain. 
 
To confirm and characterise the systems in terms of masses, bulk densities, and orbital parameters, we monitored the systems with the FIES spectrograph, and the Tull Coude Spectrograph \citep{Tull1995} at the 2.7\,m Harlan J. Smith telescope at the McDonald Observatory, Texas, USA. The FIES and Tull spectrographs are both cross-dispersed spectrographs with resolving powers of 67,000 (in high-resolution mode) and 60,000, respectively. Finally, to investigate companionship in the systems, we obtained speckle imaging using the 2.5-m reflector at the Caucasian Mountain Observatory of Sternberg Astronomical Institute \citep[CMO SAI;][]{Shatsky2020}.

\subsection{Speckle interferometry with SPP}

\begin{figure*}
    \centering
    \includegraphics[width=\textwidth]{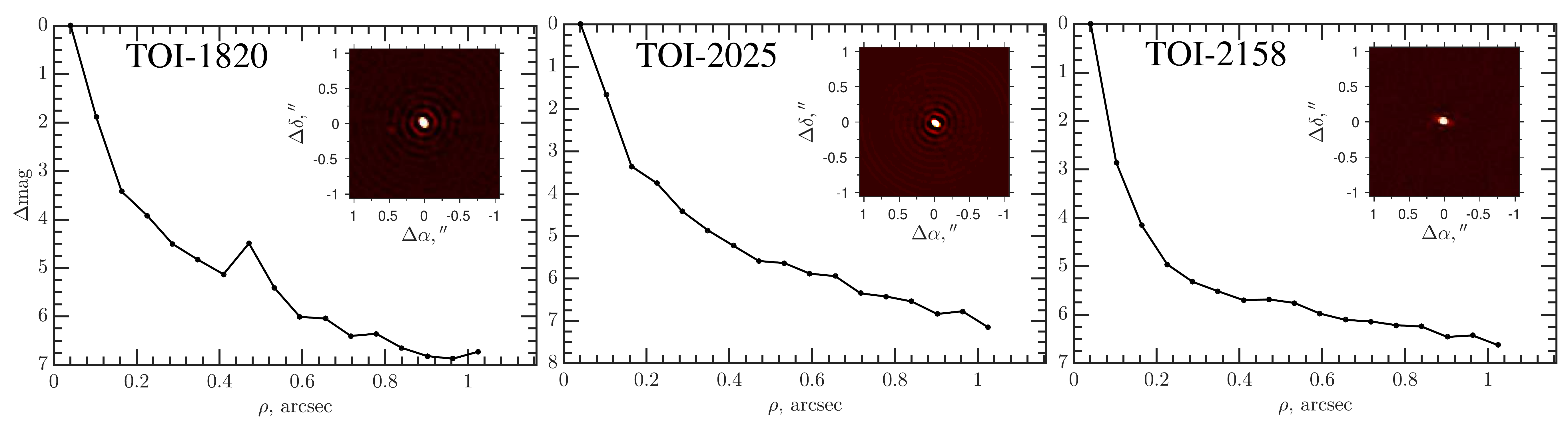}
    \caption{{\bf Speckle interferometry.} SAI-2.5m speckle sensitivity curve and autocorrelation function (ACF) for TOI-1820 (left panel), TOI-2025 (middle panel), and TOI-2158 (right panel). All images shown here were taken in the $I$-band. Only the speckle image of TOI-1820 shows evidence of a nearby companion, as can be seen by the bump in the ACF around 0.45~arcsec.}
    \label{fig:speckle}
\end{figure*}

TOI-2158, TOI-2025, and TOI-1820 were observed using speckle interferometry with the SPeckle Polarimeter (SPP; \citealt{Safonov2017}) on the 2.5-m telescope at the Sternberg Astronomical Institute of Lomonosov Moscow State University (SAI MSU). The detector has a pixel scale of 20.6~mas px$^{-1}$, and the angular resolution was 83~mas. The atmospheric dispersion compensation by two direct vision prisms allowed us to use the relatively broadband $I_c$ filter. For all targets, 4000 frames of 30~ms were obtained. The detection limits are provided in \fref{fig:speckle}. For TOI-2158 and TOI-2025, we did not detect any stellar companions, with limits for $\Delta$mag for any potential companion of 6.5~mag and 7~mag at $1^{\prime\prime}$, respectively.

\subsubsection{Stellar companion to TOI-1820}

For TOI-1820 we detected a companion 4.0 magnitudes fainter than the primary on December 2, 2020 and July 15, 2021. The separation, position angle, and contrast were determined by the approximation of the average power spectrum with the model of a binary star \citep[see Eq. (9) in][]{Safonov2017}. As the weight for the approximation, we took the inverse squared uncertainty of the power spectrum determination. The results are presented in \tref{tab:spp_TOI1820}. All binarity parameters for the two dates coincide within the uncertainties. According to \gaia EDR3 \citep{GaiaEDR3}, the proper motion of TOI-1820 is relatively high, being $50.54\pm0.08$~mas\,yr$^{-1}$ and $-33.93\pm0.08$~mas\,yr$^{-1}$ along right ascension and declination, respectively. If the companion of TOI-1820 were a background star, its position with respect to TOI-1820\footnote{In the SIMBAD entry \url{http://simbad.u-strasbg.fr/simbad/sim-basic?Ident=TYC+1991-1863-1&submit=SIMBAD+search}, TOI-1820 is listed as a member of the cluster Melotte 111. However, the proper motion ($\mu_\alpha\sim-12$~mas yr$^{-1}$, $\mu_\delta\sim-9$~mas yr$^{-1}$) and parallax ($\varpi\sim12$~mas) are significantly different from the \gaia EDR3 \citep{GaiaEDR3} values listed in \tref{tab:targets}.} would change by $37.694\pm0.051$~mas between the two epochs of our observations. As long as we see a displacement much smaller than this, we conclude that TOI-1820 and its companion are gravitationally bound. With a \gaia parallax of 4~mas (see \tref{tab:targets}), we find a physical separation between the target and the companion of $\approx$110~AU. Furthermore, from our HIRES reconnaissance and using the algorithm from \citet{Kolbl2015}, we can constrain this secondary companion to only contribute 1\% in flux if the radial velocity (RV) separation between the components in TOI-1820 is greater than 10 km/s. If the RV separation were less than 10 km~s$^{-1}$, the flux of the secondary would have been unconstrained without the speckle interferometry.

\begin{table}
   \caption{{\bf Speckle observations of TOI-1820.}}
\begin{threeparttable}
\begin{tabular}{ccccc}
\hline
Date & Separation & P.A. & $\Delta m$ \\
UT & mas & $^{\circ}$ &  \\
\hline
2020-12-02 & $470\pm5$ & $102.6\pm0.3$ & $4.0\pm0.1$ \\
2021-07-15 & $474\pm8$ & $101.7\pm0.9$ & $3.7\pm0.1$ \\
\hline
\end{tabular}
\begin{tablenotes}
    \item  Results from the SPP speckle interferometry of TOI-1820: separation, position angle, and
contrast.
\end{tablenotes}
\end{threeparttable}
   \label{tab:spp_TOI1820}
\end{table}

\subsection{Photometric follow-up}

We acquired ground-based time-series follow-up photometry of TOI-1820, TOI-2025, and TOI-2158 as part of the \textit{TESS} Follow-up Observing Program \citep[TFOP;][]{collins:2019}\footnote{https://tess.mit.edu/followup} to attempt to: (1) rule out or identify nearby eclipsing binaries (NEBs) as potential sources of the detection in the \textit{TESS} data; (2) detect the transit-like events on target to confirm the depth, and thus the \textit{TESS} photometric deblending factor; (3) refine the \textit{TESS} ephemeris; and (4) place constraints on transit depth differences across optical filter bands. We used the {\tt TESS Transit Finder}, which is a customised version of the {\tt Tapir} software package \citep{Jensen:2013}, to schedule our transit observations. Unless otherwise noted, the images were calibrated and the photometric data were extracted using the {\tt AstroImageJ} ({\tt AIJ}) software package \citep{Collins:2017}. The observing facilities are described below, and the individual observations are detailed in Table \ref{table:transitfollowup}. The ground-based light curves for TOI-1820, TOI-2025, and TOI-2158 are shown in \fref{fig:lc_toi1820}, \fref{fig:lc_toi2025}, and \fref{fig:lc_toi2158}, respectively.

We observed six transits using the Las Cumbres Observatory Global Telescope \citep[LCOGT;][]{Brown:2013} 1.0-m and 0.4-m networks. Three transits were observed in alternating filter mode, resulting in a total of nine light curves. The 1-m telescopes are equipped with $4096\times4096$ pixel SINISTRO cameras having an image scale of $0\farcs389$ per pixel, resulting in a $26\arcmin\times26\arcmin$ field of view. The 0.4-m telescopes are equipped with $2048\times3072$ pixel SBIG STX6303 cameras having an image scale of 0$\farcs$57 pixel$^{-1}$, resulting in a $19\arcmin\times29\arcmin$ field of view. The images were calibrated by the standard LCOGT {\tt BANZAI} pipeline \citep{McCully:2018}.

We observed a transit from KeplerCam on the 1.2-m telescope at the Fred Lawrence Whipple Observatory using alternating filters, resulting in two light curves. The $4096\times4096$ Fairchild CCD 486 detector has an image scale of $0\farcs336$ per pixel, resulting in a $23\farcm1\times23\farcm1$ field of view.

We observed one transit each from the Kotizarovci Private Observatory 0.3-m telescope near Viskovo, Croatia, the C.R. Chambliss Astronomical Observatory (CRCAO) 0.6-m telescope at Kutztown University near Kutztown, PA, and the Conti Private Observatory 0.3-m telescope near Annapolis, MD. The Kotizarovci telescope is equipped with a $765\times510$ pixel SBIG ST7XME camera having an image scale of $1\farcs2$ per pixel, resulting in a $15\arcmin\times10\arcmin$ field of view. The CRCAO telescope is equipped with a $3072\times2048$ pixel SBIG STXL-6303E camera having an image scale of $0\farcs76$ after $2\times2$ pixel image binning, resulting in a $13\arcmin\times20\arcmin$ field of view. The Conti telescope is equipped with a $2750\times2200$ pixel StarlightXpress SX694M camera having an image scale of $1\farcs0$ after $2\times2$ pixel image binning, resulting in a $23\arcmin\times18\arcmin$ field of view.

\subsection{RV follow-up}
Our NOT and McDonald Observatory monitoring was carried out from May 2020 to June 2022. In \tref{tab:rv_toi1820} and \tref{tab:rv_toi2025} we list all epochs and RVs for TOI-1820 and TOI-2025, respectively. \tref{tab:rv_toi2158a} and \tref{tab:rv_toi2158b} contain all epochs and RVs for and TOI-2158.

We reduced the FIES spectra using the methodology described in \citet{Buchhave2010} and \citet{Gandolfi2015}, which includes bias subtraction, flat fielding, order tracing and extraction, and wavelength calibration. We traced the RV drift of the instrument acquiring long-exposed ThAr spectra ($\sim$80~s) immediately before and after each science observation. The science exposure time was set between1800-2700~seconds, depending on the sky conditions and scheduling constraints. As our exposures were longer than 1200~s, we split the exposure in three sub-exposures to remove cosmic ray hits using a sigma clipping algorithm while combining the frames.  RVs were derived via multi-order cross-correlations, using the first stellar spectrum as a template. 

For Tull we used 30-minute integrations to give a S/N of 60-70 per pixel. An $I_2$ gas absorption was used to provide the high-precision RV metric. All Tull spectra were reduced and extracted using standard IRAF tasks. Radial velocities were extracted using the Austral code \citep{Endl2000}.

\begin{figure*}
    \centering
    \includegraphics[width=\textwidth]{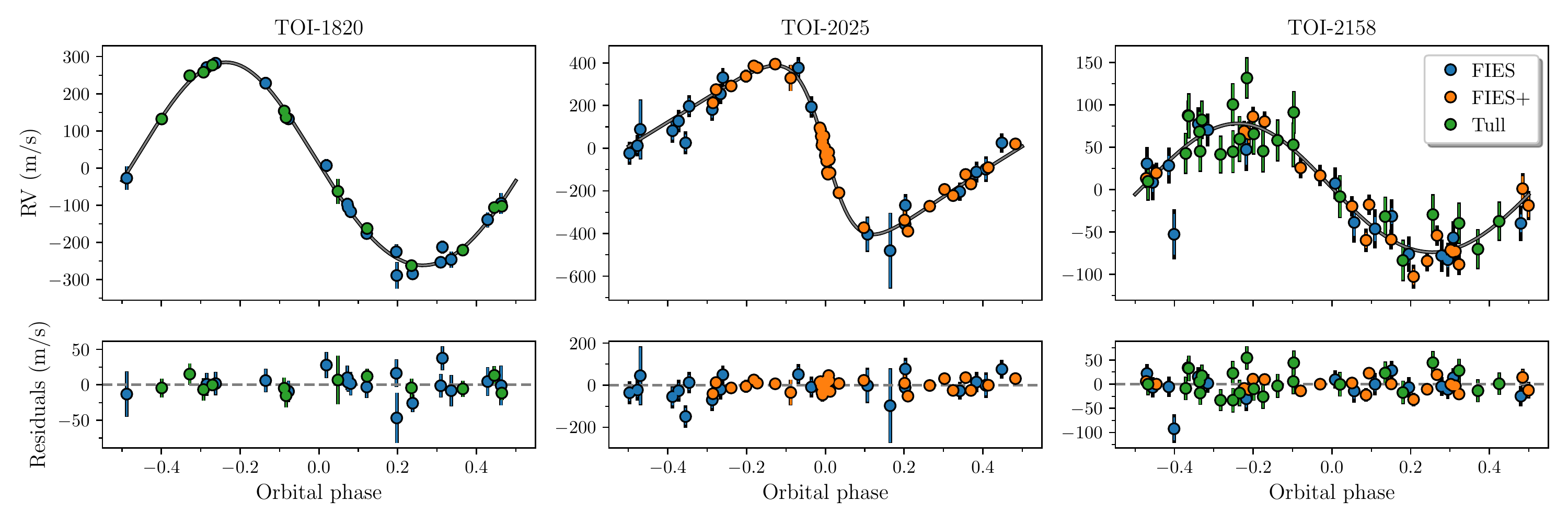}
    \caption{{\bf Radial velocities.} From left to right are our FIES (blue), FIES+ (orange), and Tull (green) RVs for TOI-1820, TOI-2025, and TOI-2158, respectively, where the black parts of the error bars denote the jitter added in quadrature. The grey curves are the best-fitting models. In the bottom row are the residuals after subtracting the best-fitting models.}
    \label{fig:rv_all}
\end{figure*}

\begin{figure*}
    \centering
    \includegraphics[width=\textwidth]{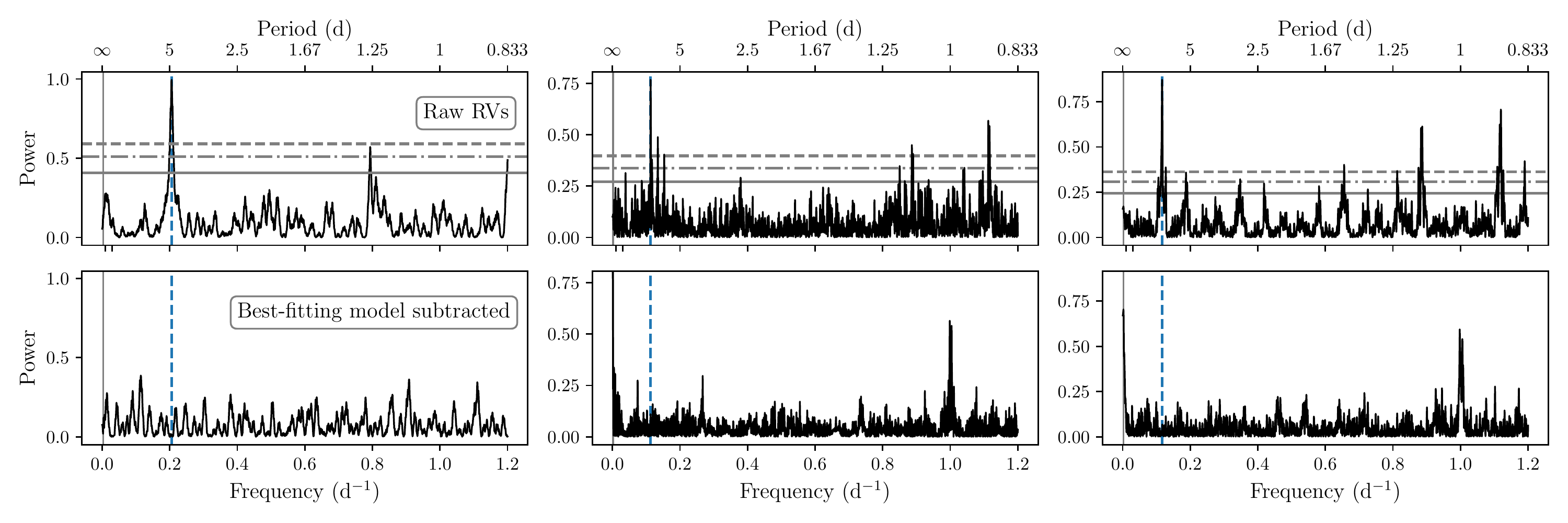}
    \caption{{\bf Generalised Lomb-Scargle periodograms}. From left to right are the GLS periodograms for TOI-1820, TOI-2025, and TOI-2158. In the top row, we show the GLS periodograms directly from the RVs, and in the bottom we have removed the orbit of the planet. The vertical dashed lines from left to right denote the 4.9~d, 8.9~d, and 8.6~d signals seen in the photometry for TOI-1820, TOI-2025, and TOI-2158, respectively. The solid lines are our baselines, i.e. $1/(t_\mathrm{last \ RV} - t_\mathrm{first \ RV})$ with $t_\mathrm{first \ RV}$ and $t_\mathrm{last \ RV}$ being the times for the first and last acquired RVs. The horizontal solid, dot-dashed, and dashed lines show the 10\%, 1\%, and 0.1\% false alarm probabilities, respectively.}
    \label{fig:gls_all}
\end{figure*}

To validate the planetary nature of the transiting signal in TOI-1820 and fully characterise the system, we acquired 18 spectra with FIES and 12 spectra with Tull, shown to the left in \fref{fig:rv_all}. \fref{fig:gls_all} displays the generalised Lomb-Scargle \citep[GLS;][]{Lomb76,Scargle82} periodograms with TOI-1820 to the left, in which the $\sim$4.9~d transiting signal has been overplotted as the dashed line. This periodicity corresponds to the peak that we see in the GLS of the RVs.

We collected a total of 46 FIES RVs to validate the planetary nature of the signal, as well as to characterise the TOI-2025 system. In the middle panel of \fref{fig:rv_all}, FIES+ refers to RVs collected after July 1, 2021 (see Section~\ref{sec:mcmc}). As before, the transiting signal coincides with the peak in the GLS periodogram in the middle panels of \fref{fig:gls_all}.

For TOI-2158 we collected 30 FIES RVs and 23 Tull RVs, shown in the right panel of \fref{fig:rv_all}. As for the other two systems, the peak associated with the $\sim$8.6~d period planet is detected in the GLS periodogram in \fref{fig:gls_all}, since it is stronger than the false alarm probability.

\section{Stellar parameters}\label{sec:stelpars}
We made use of the stellar parameter classification \citep[SPC;][]{Buchhave2012,Buchhave2014,Bieryla2021} tool to obtain stellar parameters, where we reduced and extracted the spectra following the approach in \citet{Buchhave2010}. For TOI-2025 and TOI-2158, we used the TRES spectra as reconnaissance, and for TOI-1820, we used our FIES spectra. The derived stellar parameters are tabulated in \tref{tab:targets}.

In addition, for TOI-1820 we also used our HIRES spectra with \texttt{Specmatch-Synth} to derive stellar parameters as described in \citet{Petigura2017}. From the two HIRES spectra, we find $T_\mathrm{eff}=5695 \pm 100$~K, $\log g=4.1 \pm 0.1$, [Fe/H]$=0.01\pm0.06$, and $v \sin i = 3.07 \pm 0.77$~km~s$^{-1}$. We also estimated the R$^\prime_\mathrm{HK}$ activity indicator. As a result we obtained $\log $R$^{\prime}_\mathrm{HK}$ = -5.37, a hint that the star is inactive.

\subsection{SED}
\begin{figure*}
    \centering
    \includegraphics[width=\textwidth]{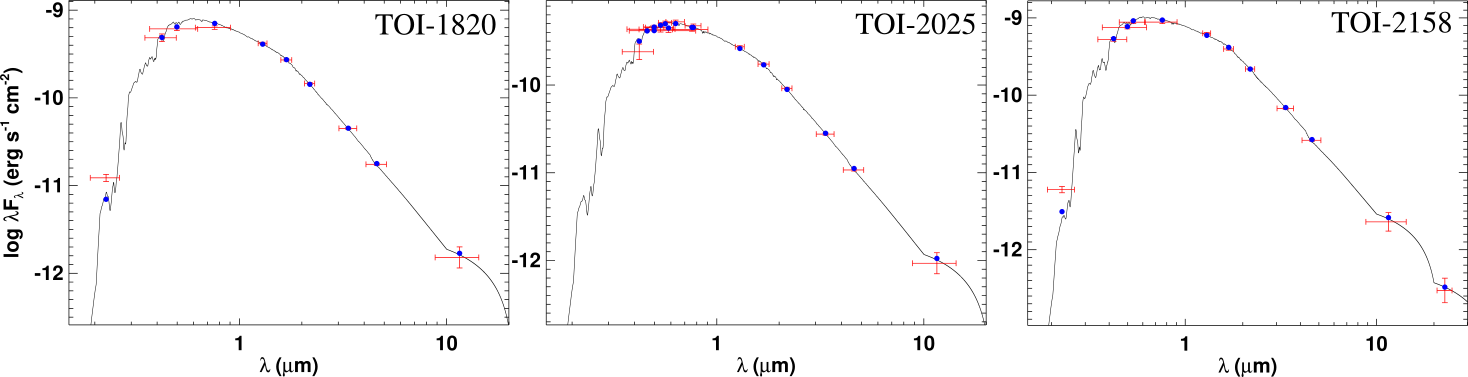}
    \caption{{\bf Spectral Energy Distribution.} The SEDs for TOI-1820 (left panel), TOI-2025 (middle panel), and TOI-2158 (right panel). Red symbols represent the observed photometric measurements, where the horizontal bars represent the effective width of the passband. Blue symbols are the model fluxes from the best-fit Kurucz atmosphere model (black).} 
    \label{fig:sed}
\end{figure*}

As an independent check on the derived stellar parameters, we performed an analysis of the broadband spectral energy distribution (SED) together with the {\it Gaia\/} EDR3 \citep{GaiaEDR3} parallax in order to determine an empirical measurement of the stellar radius, following the procedures described in \citet{Stassun:2016,Stassun:2017,Stassun:2018}. In short, we pulled the $B_T V_T$ magnitudes from Tycho-2, the $BVgri$ magnitudes from APASS, the $JHK_S$ magnitudes from {\it 2MASS}, the W1--W4 magnitudes from {\it WISE}, and the $G G_{\rm BP} G_{\rm RP}$ magnitudes from {\it Gaia}. We also used the {\it GALEX} NUV flux when available. Together, the available photometry spans the stellar SED over the wavelength range 0.35--22~$\mu$m, and extends down to 0.2~$\mu$m when {\it GALEX} data are available (see \fref{fig:sed}). We performed a fit using Kurucz stellar atmosphere models, with the priors on effective temperature ($T_{\rm eff}$), surface gravity ($\log g$), and metallicity ([Fe/H]) from the spectroscopically determined values. The remaining free parameter was the extinction ($A_V$), which we restricted to the maximum line-of-sight value from the dust maps of \citet{Schlegel:1998}. 

The resulting SED fits are shown in \fref{fig:sed} for TOI-1820, TOI-2025, and TOI-2158 with reduced $\chi^2$ values of 1.5, 1.2, and 1.2, respectively. The resulting best-fit are summarised in \tref{tab:targets}. Integrating the (unreddened) model SED gives the bolometric flux at Earth, $F_{\rm bol}$, which with the $T_{\rm eff}$ and the {\it Gaia\/} EDR3 parallax \citep[with no systematic adjustment; see][]{StassunTorres:2021} gives the stellar radius. The stellar mass can then be determined empirically from the stellar radius and the spectroscopic $\log g$, and compared to the mass estimated from the empirical relations of \citet{Torres:2010}. Finally, we can estimate the age of the star from the spectroscopic $R'_{\rm HK}$ via the empirical relations of \citet{Mamajek:2008}, which we can also corroborate by comparing the stellar rotation period predicted at that age from the empirical gyrochronology relations of \citet{Mamajek:2008} against that determined from the stellar radius together with the specroscopic $v\sin i$. These parameters are also summarised in \tref{tab:targets}. The rather old ages inferred for TOI-1820 and TOI-2158 would predict slow stellar rotation periods of $P_{\rm rot} = 40 \pm 2$~d and $P_{\rm rot} = 43 \pm 3$~d, respectively, whereas the (projected) rotational periods estimated from the spectroscopic $v\sin i$ together with $R_\star$ gives $P_{\rm rot} / \sin i = 24.9 \pm 6.3$~d and $P_{\rm rot} / \sin i = 19.3 \pm 3.2$~d, suggesting either somewhat younger ages, or  a process that kept the stars rotating faster than expected for their ages.

It is interesting that both TOI-1820 and TOI-2158 appear to be rotating faster than what would be expected given their ages, especially seeing as both of these stars host a hot Jupiter. Discrepancy between ages inferred from isochrone fitting and gyrochronology among hot Jupiter hosts has been seen in studies by \citet{Brown2014} and \citet{Maxted2015}, and both studies suggested tidal spin-up as a possible explanation. Further evidence for this has recently been found in \citet{Arevalo2021}. Tidal spin-up might, therefore, be the mechanism responsible for the discrepancy we are seeing in TOI-1820 and TOI-2158. Of course, this might also apply to the TOI-2025 system as this system also harbours a hot Jupiter, but as this system is younger, the effect might be less pronounced. We examined the residuals of the light curves from our best-fitting models (\fref{fig:lc_tess2min}) to see if we could see any signs of stellar variability, for instance, rotation. However, we did not detect any signals.

\section{Joint analysis}
\label{sec:mcmc}
To estimate the planetary and orbital parameters, we fit the photometry and the RVs jointly, where we extracted confidence intervals through Monte Carlo Markov chain (MCMC) sampling using the \texttt{emcee} package by \citet{Foreman}. We modelled the light curves using the \texttt{batman} package \citep{Kreidberg}, which utilises the formalism by \citet{Mandel2002}. To account for any morphological light curve distortion \citep{Kipping2010} caused by the 30-minute sampling, we oversampled our 30-minute-cadence light curves to correspond to a sampling of 2~minutes.

In an attempt to mitigate correlated noise in the \tess photometry, we made use of Gaussian process (GP) regression through the \texttt{celerite} package \citep{celerite}. We used the Matérn-3/2 kernel, which includes two hyper parameters: the amplitude of the noise, $A$, and the timescale, $\tau$. The only correction to the \tess data prior to the MCMC was the aforementioned background correction. For our ground-based photometry, we did not have long out-of-transit baselines. Therefore, we did not model the noise from these transits with GPs, instead we used a Savitsky-Golay filter to de-trend the data with each draw in our MCMC.

To fit the RVs we used a Keplerian orbit, where we naturally had different systemic velocities, $\gamma$, for the RVs stemming from FIES and Tull, when this is relevant. Due to a refurbishment of the FIES spectrograph, an offset in RV was introduced between the RVs obtained before July 1, 2021 and those obtained after. We assigned two independent systemic velocities and two independent jitter terms to RVs obtained before (FIES) and after (FIES+) this date.

Our MCMC analysis for the three systems stepped in $\cos i$ instead of $i$, as well as in $\sqrt{e}\cos \omega$ and $\sqrt{e}\sin \omega$ instead of $e$ and $\omega$. Furthermore, the code stepped in the sum of the limb darkening parameters, namely $q_1 + q_2$, where we applied a Gaussian prior with a width of 0.1. We instead fixed the difference fixed, $q_1 - q_2$, during the sampling. We retrieved the starting values of $q_1$ and $q_2$ for the \tess passband from the table \citet{Claret17}, while we used the values from \citet{Claret2013} for the ground-based photometry. Furthermore, we used $V$ as a proxy for our transit observations of TOI-2025 using FIES. The initial and resulting values for the limb-darkening coefficients can be found in \tref{tab:ld_toi1820}, \tref{tab:ld_toi2025}, and \tref{tab:ld_toi2158}.

We list all the adopted priors in \tref{tab:priors}, where a hyphen denotes that the associated parameter is not relevant for that run. We define our likelihood function as
\begin{equation}
    \label{equ:likelihood}
    \log \mathcal{L} =-0.5 \sum_{i=1}^{N} \left [ \frac{(O_i - C_i)^2}{\sigma_i^2} + \log 2 \pi \sigma_i^2 \right] + \sum_{j=1}^{M} \log \mathcal{P}_{j}\, ,
\end{equation}
where $N$ indicates the total number of data points from photometry and RVs. $C_i$ represents the model corresponding to the observed data point $O_i$. $\sigma_i$ represents the uncertainty for the $i$th data point, where we add a jitter term in quadrature and a penalty in the likelihood for the RVs. $\mathcal{P}_j$ is the prior on the $j$th parameter.

We ran our MCMC until convergence, which we assessed by looking at the rank-normalised $\hat{R}$ diagnostic test as implemented in the \texttt{rhat} module in \texttt{ArviZ} \citep{arviz_2019}.

\subsection{TOI-1820}
Given the large separation of around 110~AU for the companion, the orbital period must be rather large and the expected $K$-amplitude must be rather small, meaning that, even if it is bound, it will not affect our RVs. The companion will, however, dilute the light curve. We therefore include a contaminating factor, where we write the total flux as a function of time as $F(t)=(F_1(t) + F_2)/(F_1 + F_2)$ with $F_1(t)$ and $F_1$ being the flux respectively in- and out-of-transit from the planet hosting star, and $F_2$ is the (constant) flux from the contaminating source (or sources). Here, we included the flux from the contaminating source as a fraction of the host, $F_2/F_1$, as the difference in magnitude, namely $\delta \rm M = -2.5 \log (F_2/F_1)$. Conveniently, $\delta \rm M$ is derived from observations in the $I$-band, which is close to the bandpasses from \tess, $i^\prime$, and $z$-short (\fref{fig:lc_toi1820}). However, the dilution might be overestimated in the $B$-band. Therefore, we adopted a different value for the $B$-band of $\delta M=4.5 \pm 1.0$ as the companion is most likely a cooler star.

\subsection{TOI-2025}
\begin{figure}
    \centering
    \includegraphics[width=\columnwidth]{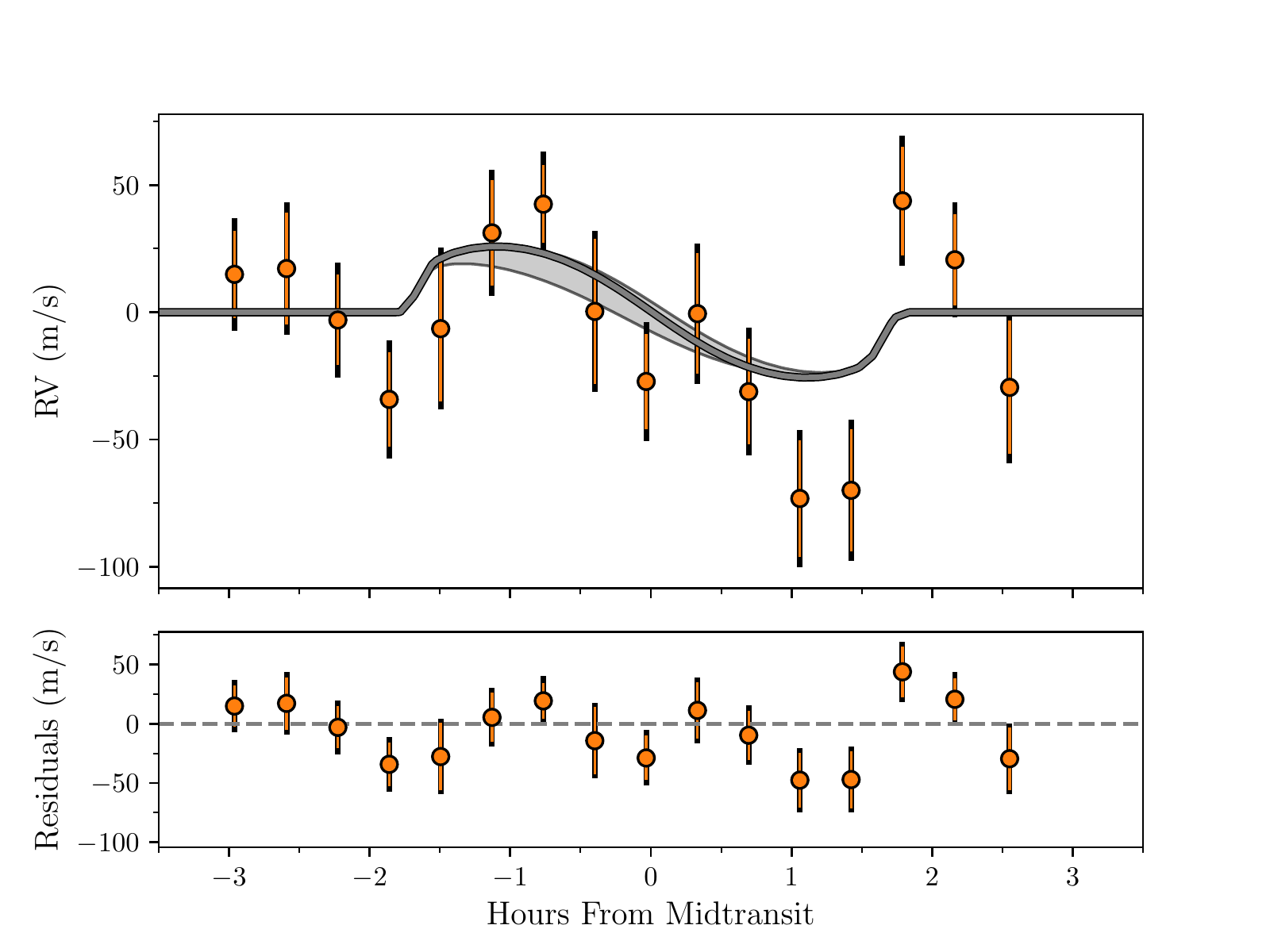}
    \caption{{\bf Rossiter-McLaughlin effect in TOI-2025.} Our in-transit observations of TOI-2025 with FIES+. {\it Top:} The Keplerian orbit and quadratic trend has been subtracted from the RVs to  better show the RM effect, with the grey line being the best-fitting model. The shaded area denotes the confidence interval in the projected obliquity, $\lambda=$\lamtwenty. {\it Bottom:} Here we have further subtracted this best-fitting model from the RVs.}
    \label{fig:rm_toi2025}
\end{figure}

For TOI-2025, we have two sets of light curves with different cadences (2~min. and 30~min.), and we apply two different oversampling factors, while using the same limb darkening coefficients for both. We observed a spectroscopic transit of TOI-2025 at the NOT (FIES+) on the night starting on the August 8, 2021, allowing us to determine the projected obliquity, $\lambda$, of the host star. The RVs obtained during this transit night can be seen in \fref{fig:rm_toi2025}. We therefore also included a model for the Rossiter-McLaughlin \citep[RM;][]{Rossiter,McLaughlin} effect using the algorithm by \citet{Hirano2011} for this fit. We used our SPC value in \tref{tab:targets} for $v \sin i_\star$ as a prior. For the macro- and micro-turbulence, we used priors stemming from the relations in \citet{Doyle2014} and \citet{Bruntt10}, respectively, along with the stellar parameters in \tref{tab:targets}.

\begin{figure}
    \centering
    \includegraphics[width=\columnwidth]{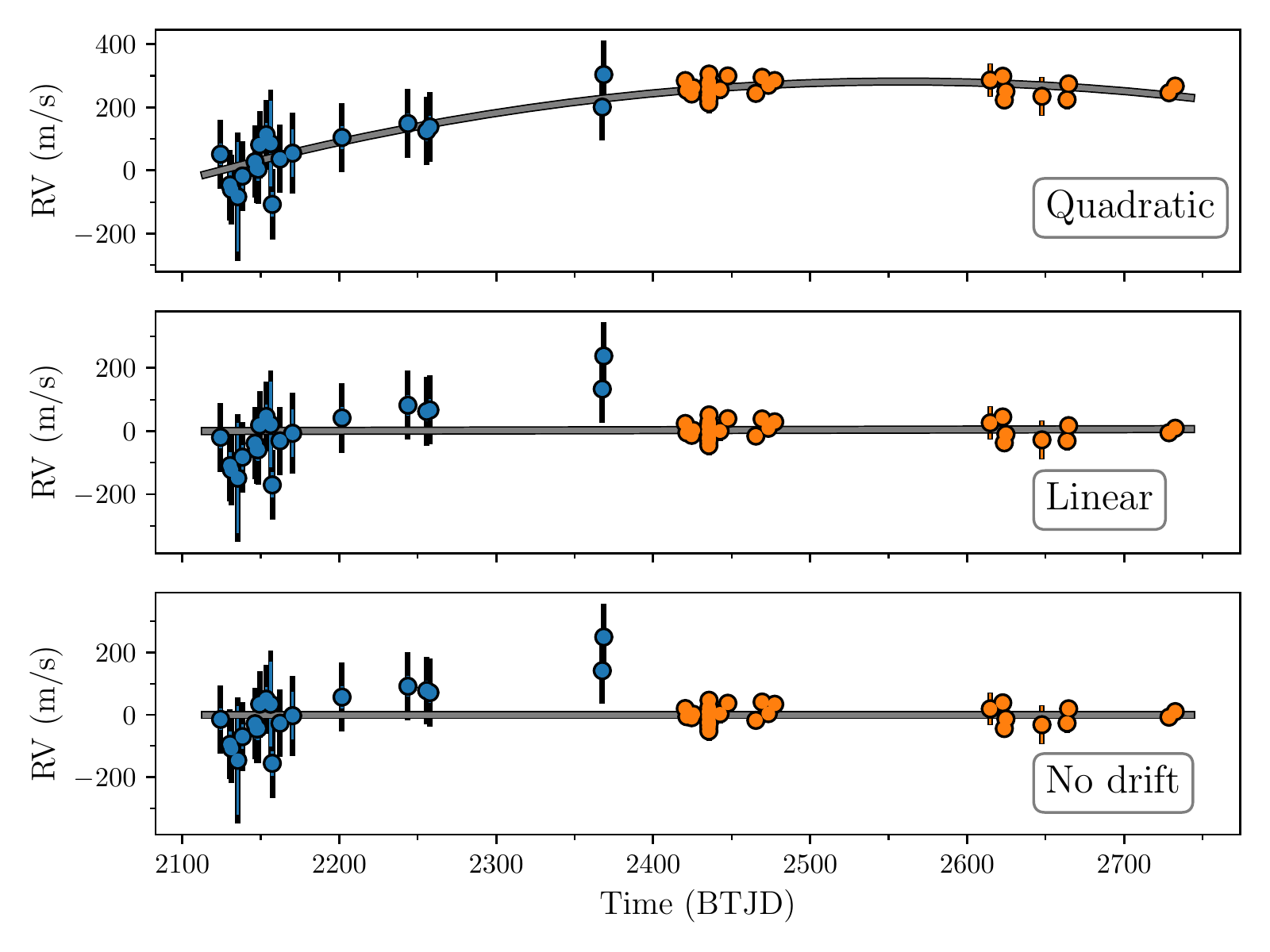}
    \caption{{\bf Long-term trend in TOI-2025.} The symbols are the same as in \fref{fig:rv_all}, but here the RVs are plotted against time, and we have subtracted the planetary signal. {\it Top:} A fit where we allow for a quadratic trend. {\it Middle:} A fit where we only allow for a linear trend. {\it Bottom:} Here we do not include any long-term drift.}
    \label{fig:rv_toi2025_drift}
\end{figure}

We carried out three MCMC runs for TOI-2025 to investigate the long-term trend: 1) a run where we included two additional parameters: a second order, $\ddot{\gamma}$, and a first-order acceleration parameter, $\dot{\gamma}$; 2) a run where we only included the first order parameter; and 3) a run where we did not allow for any long-term drift. These three runs are shown in \fref{fig:rv_toi2025_drift}. 

\subsection{TOI-2158}

\begin{figure}
    \centering
    \includegraphics[width=\columnwidth]{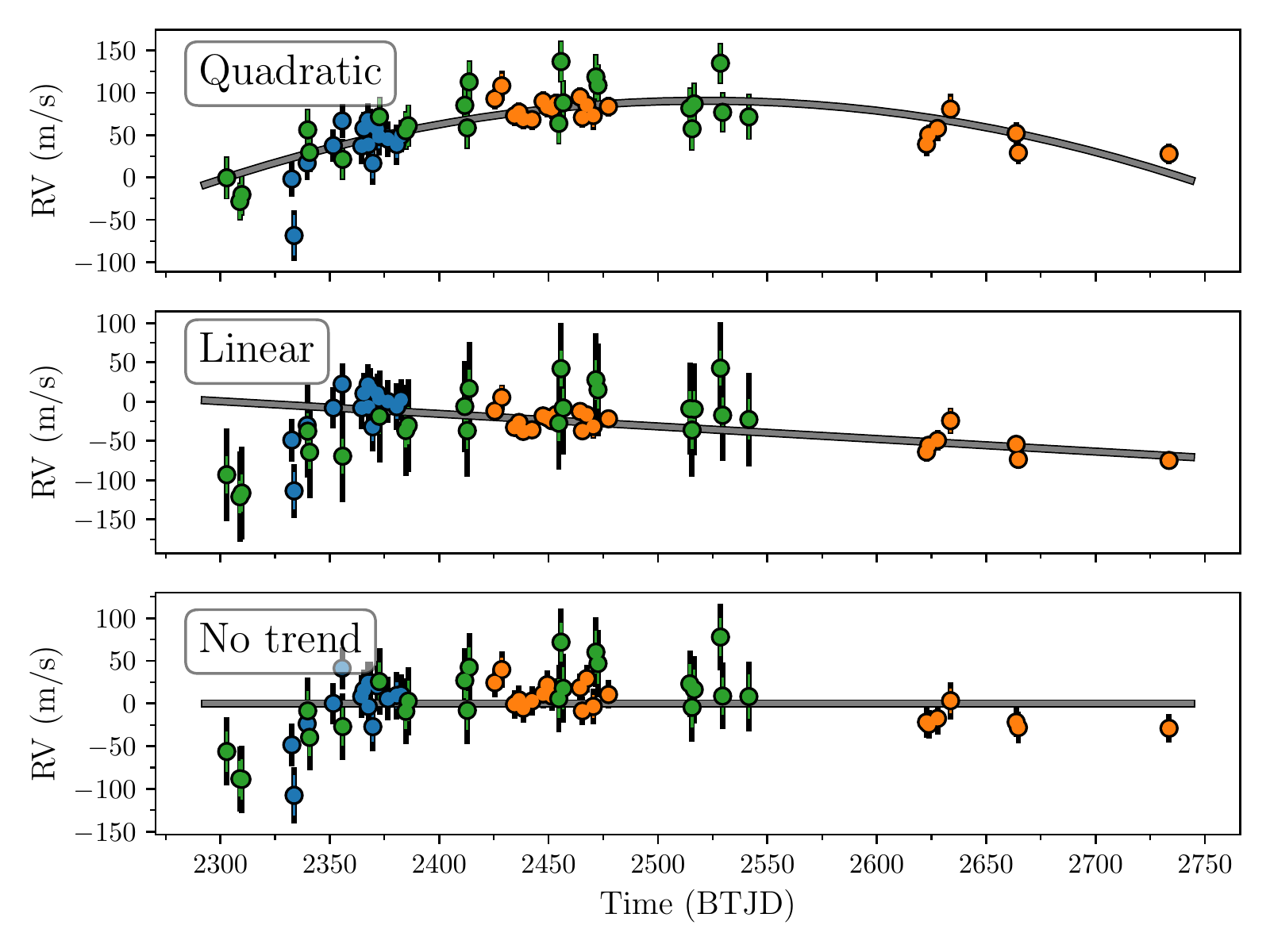}
    \caption{{\bf Long-term trend in TOI-2158.} The symbols are the same as in \fref{fig:rv_all}, but here the RVs are plotted against time, and we have subtracted the planetary signal. {\it Top:} A fit where we allow for a quadratic trend. {\it Middle:} Here we only allow for a linear trend. {\it Bottom:} Here we do not include any long-term drift.}
    \label{fig:rv_toi2158_drift}
\end{figure}

Similarly to the case of TOI-2025, the RVs of TOI-2158 show a long-term trend. We therefore performed the same three runs as for TOI-2025. These are shown in \fref{fig:rv_toi2158_drift}.

\section{Results}
\label{sec:results}
The results from the MCMC for our preferred orbital configuration for each of the systems are tabulated in \tref{tab:mcmc}. We find that TOI-1820b is a Jupiter-sized planet, \radeighteen, but significantly more massive, \masseighteen. With an orbital period of \pereighteen, it is the planet with the shortest orbital period in our sample. TOI-2025 has a similar size, \radtwenty, as TOI-1820, but has about twice its mass, \masstwenty. On the other end of the mass spectrum, we find TOI-2158~b with \masstwentyone. TOI-2158~b is also a bit smaller than the two other planets with a radius of \radtwentyone.

\begin{table*}[t]
    \centering
    \caption{{\bf Results from our MCMC analysis.}}
    \begin{threeparttable}
    \begin{tabular}{l l c c c c}
    \toprule
         \multicolumn{2}{l}{Parameter} & TOI-1820 & TOI-2025 & TOI-2158 \\ \midrule
$P$ & Period (days) & $4.860674\pm0.000005$ & $8.872078\pm0.000007$ & $8.60077\pm0.00003$ \\ 
 $T_0$ & Mid-transit time (BJD) & $2458903.0638\pm0.0006$ & $2458690.2898\pm0.0004$ & $2459018.9225_{-0.0011}^{+0.0010}$ \\ 
 $R_\mathrm{p}/R_\star$ & Planet-to-star radius ratio & $0.0777\pm0.0009$ & $0.0736\pm0.0004$ & $0.0700\pm0.0009$ \\ 
 $a/R_\star$ & Semi-major axis to star radius ratio & $8.7\pm0.3$ & $12.7_{-0.4}^{+0.5}$ & $11.4_{-0.5}^{+0.6}$ \\ 
 $K$ & Velocity semi-amplitude (m s$^{-1}$) & $273\pm4$ & $396\pm10$ & $75\pm4$ \\ 
 $\cos i$ & Cosine of inclination & $0.097\pm0.005$ & $0.023_{-0.023}^{+0.010}$ & $0.075_{-0.006}^{+0.005}$ \\ 
 $\sqrt{e} \cos \omega$ &   & $0.20\pm0.02$ & $-0.03_{-0.02}^{+0.03}$ & $0.10_{-0.08}^{+0.10}$ \\ 
 $\sqrt{e} \sin \omega$ &   & $0.031_{-0.031}^{+0.016}$ & $0.643\pm0.016$ & $0.10_{-0.10}^{+0.05}$ \\ 
 $\gamma_1$ & Systemic velocity FIES (m s$^{-1}$) & $227_{-4}^{+5}$ & $-383\pm19$ & $13\pm9$ \\ 
 $\gamma_2$ & Systemic velocity FIES+ (m s$^{-1}$) & - & $-75_{-47}^{+45}$ & $-23_{-15}^{+14}$ \\ 
 $\gamma_3$ & Systemic velocity Tull (m s$^{-1}$) & $13947\pm4$ & - & $-64794_{-13}^{+11}$ \\ 
 $\sigma_1$ & Jitter FIES (m s$^{-1}$) & $7_{-7}^{+3}$ & $45_{-14}^{+12}$ & $19_{-7}^{+5}$ \\ 
 $\sigma_2$ & Jitter FIES+ (m s$^{-1}$) & - & $19\pm6$ & $15_{-4}^{+3}$ \\ 
 $\sigma_3$ & Jitter Tull (m s$^{-1}$) & $5_{-5}^{+2}$ & - & $12_{-11}^{+5}$ \\ 
 $\log A_1$ & GP amplitude \tess 30 min. & $-6.98_{-0.10}^{+0.09}$ & $-8.30\pm0.06$ & $-8.95_{-0.13}^{+0.12}$ \\ 
 $\log \tau_1$ & GP timescale \tess 30 min. ($\log$ days) &  $-0.77_{-0.14}^{+0.13}$ & $-0.31_{-0.11}^{+0.10}$ & $-1.7_{-0.4}^{+0.5}$ \\ 
 $\log A_2$ & GP amplitude \tess 2 min. &  $-7.36_{-0.11}^{+0.10}$ & $-7.85_{-0.10}^{+0.09}$ & $-7.271_{-0.016}^{+0.017}$ \\ 
 $\log \tau_2$ & GP timescale \tess 2 min ($\log$ days) &  $-1.03 \pm 0.14$ & $-0.23_{-0.14}^{+0.13}$ & $-7.29_{-0.06}^{+0.07}$ \\ 
 $\ddot{\gamma}$\tnote{a,b} & Quadratic trend (m s$^{-1}$ d$^{-2}$) & - & $-0.0015\pm0.0003$ & $-0.0020\pm0.0003$ \\ 
 $\dot{\gamma}$\tnote{a,b} & Linear trend (m s$^{-1}$ d$^{-1}$) & - & $1.4\pm0.2$ & $0.87_{-0.13}^{+0.14}$ \\ 
 $\delta \mathrm{M}_I$ & Dilution $I$-band/\tess & $3.9_{-0.5}^{+0.4}$ & - & - \\ 
 $\delta \mathrm{M}_B$ & Dilution $B$-band & $4.7_{-0.9}^{+0.8}$ & - & - \\ 
 $\lambda$ & Projected obliquity ($^{\circ}$) & - & $9_{-31}^{+33}$ & - \\ 
 $v \sin i_\star$ & Projected rotational velocity (km s$^{-1}$) & - & $6.0\pm0.3$ & - \\ 
 $\zeta$ & Macro-turbulence (km s$^{-1}$) & - & $4\pm1$ & - \\ 
 $\xi$ & Micro-turbulence (km s$^{-1}$) & - & $1.3_{-0.9}^{+0.7}$ & - \\ 
 \hdashline
$e$ & Eccentricity & $0.043\pm0.008$ & $0.41\pm0.02$ & $<0.070$ at $3 \sigma$\tnote{c} \\ 
 $\omega$ & Argument of periastron ($^\circ$) & $9_{-9}^{+4}$ & $93\pm2$ & $52_{-52}^{+19}$ \\ 
 $i$ & Inclination ($^\circ$) & $84.4\pm0.3$ & $88.7_{-0.6}^{+1.3}$ & $85.7_{-0.3}^{+0.4}$ \\ 
 $b$ & Impact parameter & $0.840_{-0.013}^{+0.015}$ & $0.29_{-0.29}^{+0.12}$ & $0.86_{-0.03}^{+0.02}$ \\ 
 $T \rm _{4,1}$ & Total transit duration (hours) & $2.92\pm0.04$ & $3.617_{-0.022}^{+0.017}$ & $3.77_{-0.06}^{+0.05}$ \\ 
 $T \rm _{2,1}$ & Time from 1st to 2nd contact (hours) & $0.61\pm0.05$ & $0.255_{-0.010}^{+0.009}$ & $0.75\pm0.07$ \\ 
 $R_\mathrm{p}$ & Planet radius ($\rm R_J$) & $1.14\pm0.02$ & $1.117\pm0.009$ & $0.960\pm0.012$ \\ 
 $M_\mathrm{p}$\tnote{d} & Planet mass ($\rm M_J$) & $2.3\pm0.2$ & $4.4\pm0.3$ & $0.82\pm0.08$ \\ 
 $\rho_\mathrm{p}$ & Planet density (g~cm$^{-3}$) & $2.0\pm0.2$ & $3.9\pm0.3$ & $1.14\pm0.12$ \\ 
 $T_\mathrm{eq}$\tnote{e} & Equilibrium temperature (K)\tnote{c} & $1375\pm12$ & $1167\pm11$ & $1188\pm10$ \\ 
 $a$ & Semi-major axis (AU) & $0.061\pm0.003$ & $0.092\pm0.004$ & $0.075\pm0.004$ \\ 
 
    \bottomrule     
    \end{tabular}
    
\begin{tablenotes}
    \item The parameters above the dashed line are the stepping parameters, and below are the derived parameters. The value given is the median and the uncertainty is the highest posterior density at a confidence level of 0.68.
    \item[a] Zero-point for TOI-2158 is 2459302.92570 BJD$_\mathrm{TDB}$.
    \item[b] Zero-point for TOI-2025 is 2459124.41436 BJD$_\mathrm{TDB}$.
    \item[c] Two-sided $1 \sigma$ distribution $e=0.031_{-0.031}^{+0.013}$.
    \item[d] Calculated from \eref{eq:mass}.
    \item[e] Following \citet{Kempton2018}.
\end{tablenotes}
\end{threeparttable}    
    \label{tab:mcmc}
\end{table*}

For TOI-2025 and TOI-2158, we found evidence for long-term RV trends, as can be seen in  \fref{fig:rv_toi2025_drift} and \fref{fig:rv_toi2158_drift}. In both we also saw evidence for a curvature in the RVs, which we model with a quadratic term. There is no significant evidence for long-term RV changes in TOI-1820.

Assuming the long-term RV changes are due to further-out companions, we can glimpse information about their masses from some back-of-the-envelope calculations. We can therefore obtain an order of magnitude estimate for the periods of the outer companions as $P=-2 \ddot{\gamma}/\dot{\gamma}$, resulting in periods of around 1870~d and 650~d for TOI-2025 and TOI-2158, respectively. Using the relation $K=\ddot{\gamma}P^2/4\pi^2$ derived in \citet{Kipping2011} with 

\begin{equation}
    \frac{M_{\rm p} \sin i}{\mathrm{M}_{\rm J}} = \frac{K \sqrt{1 - e^2}}{28.4~\mathrm{m}~\mathrm{s}^{-1}} \left ( \frac{P}{1~\mathrm{yr}} \right)^{1/3} \left ( \frac{M_\star}{\mathrm{M}_\odot} \right)^{2/3}  \, ,
    \label{eq:mass}
\end{equation}
we can get an estimate of the masses of the companions. From this we get masses of $\approx70$~M$_\mathrm{J}$ and $\approx15$~M$_\mathrm{J}$ for the companions in TOI-2025 and TOI-2158, respectively.

\subsection{The eccentricities of TOI-2025~b and TOI-1820~b}
We find TOI-2025~b to travel on an eccentric orbit, \ecctwenty. However, the argument of periastron is close to and fully consistent with $90^\circ$. This configuration can be deceptive when it comes to determining the eccentricity \citep[e.g.][]{Laughlin2005}. This is because the RV curves would be symmetric for values close to $|\omega|=90$~$^\circ$, even for eccentric orbits.

To further investigate the orbital eccentricity, we carried out a few experiments. First, as mentioned, we ran an MCMC where we fixed $e$ to 0. The best-fitting model from this run can be seen in \fref{fig:rv_toi2025_no_ecc}, where the residuals clearly have structure in them. Our model involving a circular orbit does apparently not capture all the complexity present in the data. Consequently, the derived RV jitter terms for both FIES and FIES+ are significantly higher, with values of $111^{+18}_{-22}$~m~s$^{-1}$ and  $82^{+10}_{-13}$~m~s$^{-1}$, respectively, as opposed to the values of \jitfiestwenty and \jitfiesptwenty from the eccentric fit. As we find a modest eccentricity for TOI-1820, we carried out a similar run for TOI-1820, finding marginally higher jitter (a couple of m~s$^{-1}$) for the $e=0$ case. 

As there might be stellar signals that are coherent on timescales of hours, but not days, and given that we have a much higher sampling during the transit night, it is worthwhile investigating if the eccentricity hinges on those measurements and to what extent. Therefore, we performed a fit in which the eccentricity was allowed to vary, but where we only included the first and the last data point from the transit night. Here, we obviously did not try to fit the obliquity. From this we get values of $e=0.42\pm0.02$ and $\omega=91 \pm 3$~$^\circ$, consistent with the values from the run using all the RV data.

\begin{figure}[t!]
    \centering
    \includegraphics[width=\columnwidth]{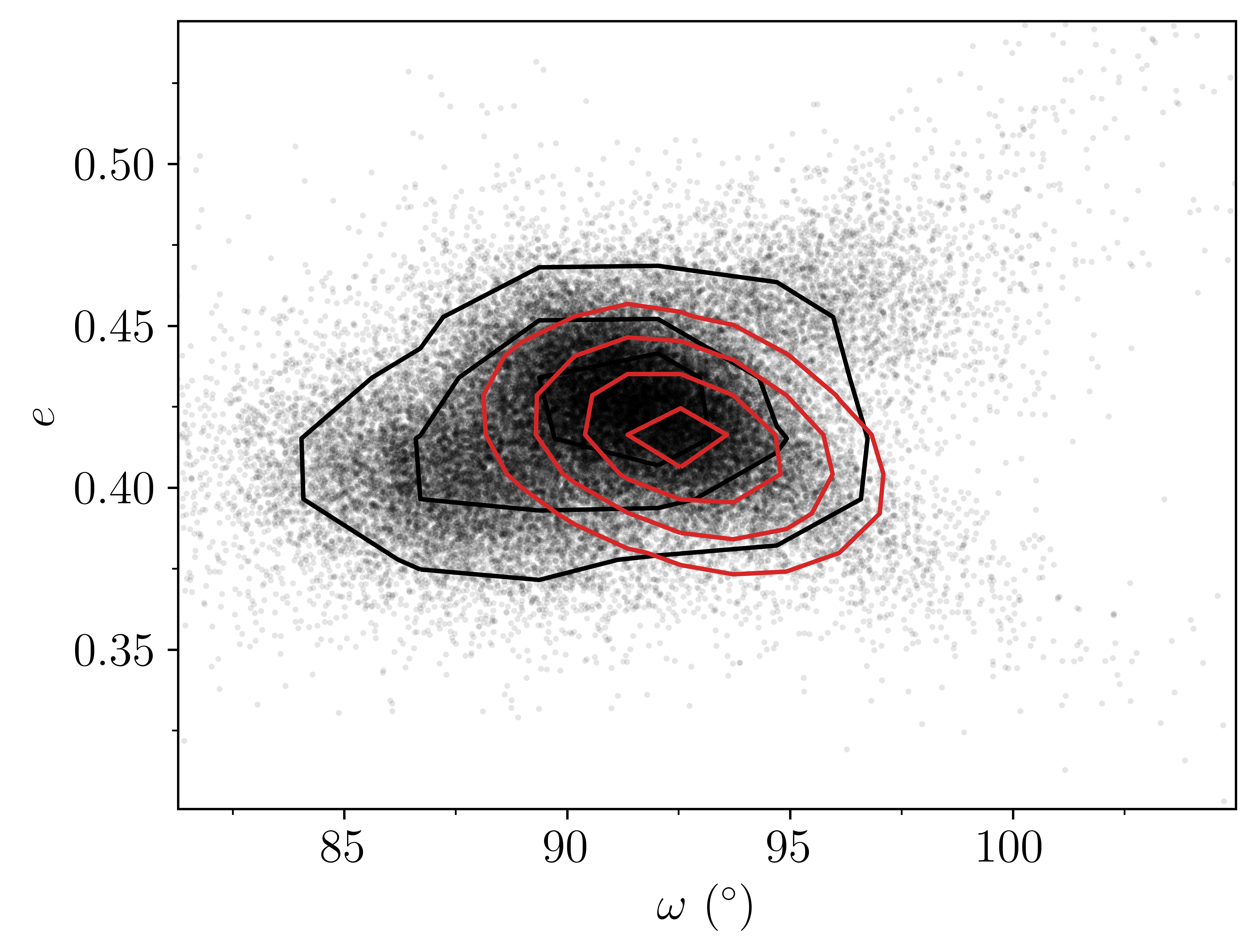}
    \caption{{\bf Bootstrapping the orbit of TOI-2025~b.} A 2D histogram of our bootstrap with 50,000 iterations displaying the eccentricity plotted against the argument of periastron. Each point is one iteration and the black contours denote the confidence levels. The red contours are the posteriors from our MCMC tabulated in \tref{tab:mcmc}.}
    \label{fig:bootstrap}
\end{figure}

\begin{figure}
    \centering
    \includegraphics[width=\columnwidth]{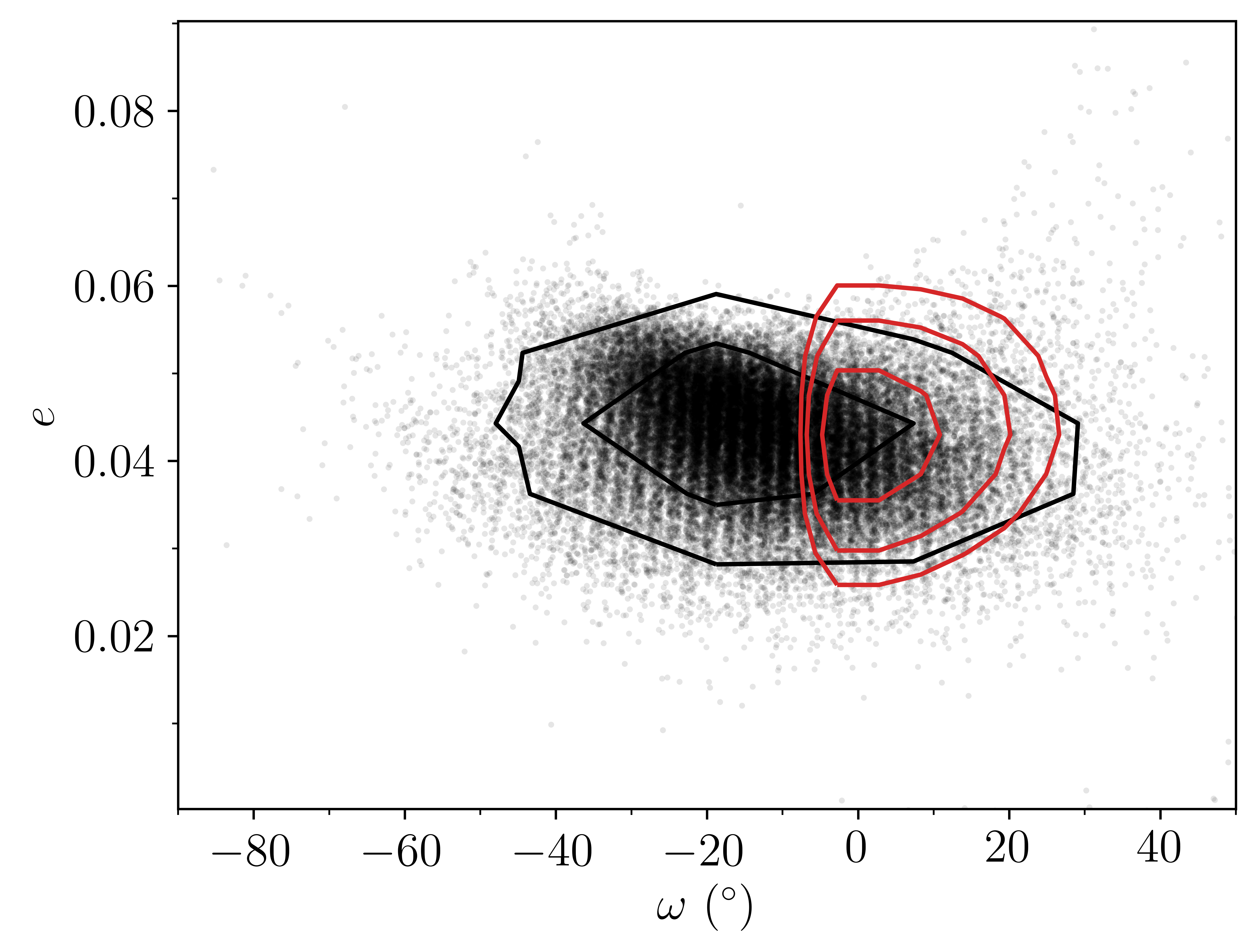}
    \caption{{\bf Bootstrapping the orbit of TOI-1820~b.} Same as in \fref{fig:bootstrap}, but for TOI-1820~b.}
    \label{fig:bootstrap_toi1820}
\end{figure}

Next we performed a bootstrap experiment using the RV data only. In our bootstrap we used alternate realisations of the data in \tref{tab:rv_toi2025}, again excluding all but the first and last data point from the transit night. After redrawing a data set from the original data, we fit for $e$, $\omega$, $\gamma_\mathrm{FIES}$, $\gamma_\mathrm{FIES+}$, $K$, $\dot{\gamma}$, and $\ddot{\gamma}$. In \fref{fig:bootstrap} we plot the results for $e$ and $\omega$ for the 50,000 realisations. Evidently, we recover an eccentric orbit even when we leave out certain data points. Therefore, we conclude that our result for the eccentricity is significant and does not hinge on a few data points. Again, we did a similar exercise for TOI-1820, which also yielded consistent results with the run from the MCMC, as seen in \fref{fig:bootstrap_toi1820}. We thus conclude that the eccentricities for TOI-2025~b and TOI-1820~b are significant (at a confidence level of $20\sigma$ and $5\sigma$, respectively), while TOI-2158~b is consistent with a circular orbit.

\subsection{The obliquity of TOI-2025}
In addition to finding an eccentric orbit for the planet, we also measured the projected obliquity of TOI-2025. We find the projected obliquity to be consistent with no misalignment, $\lambda=$\lamtwenty. The relevant transit RVs and our best-fitting model can be seen in \fref{fig:rm_toi2025}. Despite having only measured the projected obliquity, $\lambda$, here, we can make a strong argument that it is close to the obliquity, $\psi$, which requires the stellar inclination along the line of sight to be close to $90^\circ$. That $i_\star$ is close to $90^\circ$ is supported by Figure~3 in \citet{Louden2021}, where a correlation between $T_\mathrm{eff}$ and $v \sin i_\star$ is plotted. From this plot we should not expect $v \sin i_\star$ to be markedly different from the value of $6.0 \pm 0.3$~km~s$^{-1}$ given the effective temperature for TOI-2025 of $\sim5900$~K that we have found. This therefore suggests that the system is aligned.

\section{Discussion and conclusions}
\label{sec:conclusions}

We validated and characterised three hot Jupiters discovered by \tess: TOI-1820~b, TOI-2025~b, and TOI-2158~b. A\ commonality for all three systems is that we, in some way or another, see evidence for companions. The outer companions may have played a role in the migration of the gas giants, thus shaping the final architecture of the systems. \citet{Ngo2016} argue that sites hosting outer stellar companions are either more favourable environments for gas giant formation at all separations, or the presence of stellar companions might drive the inwards migration, such as through Kozai-Lidov \citep{Kozai1962,Lidov1962}, or other dynamical processes. Through our speckle interferometry of TOI-1820, we detected a $\sim$4~mag fainter stellar companion at a distance of $\sim$110~AU from the bright host. It would be interesting to obtain good estimates of the stellar parameters for this companion in order to assess whether it would have been able to drive Kozai-Lidov cycles responsible for the migration. 

If the outer companions are planets within $\sim$1~AU from the stellar host, \citet{Becker2017} found that they should be coplanar with the inner hot Jupiters, suggesting that Kozai-Lidov migration would not be viable. However, if these companions are found at greater distances (gas giants $\gtrsim$5~AU or stellar $\gtrsim$100~AU), they could still be inclined and the formation of the hot Jupiter could take place through Kozai-Lidov migration \citep{Lai2018}. In the RVs for both TOI-2025 and TOI-2158, we see long-term quadratic trends. In contrast to TOI-1820, the companions in TOI-2025 and TOI-2158 might be of planetary, or at least substellar, nature and closer in (cf. the mass and period estimates in Section~\ref{sec:results}). As the companions in TOI-2025 and TOI-2158 are most likely found beyond 1~AU, given the (lower) estimates for their periods and the stellar masses, Kozai-Lidov migration could be a viable transport mechanism for TOI-2025~b and TOI-2158~b. \tess might be able to shed more light on these outer companions as more sectors become available. According to the Web \tess Viewing Tool\footnote{\url{https://heasarc.gsfc.nasa.gov/cgi-bin/tess/webtess/wtv.py}}, TOI-2025 should be observed again in Sectors 52, 53, and 58-60, and TOI-2158 is set to be observed in Sector 53. In addition, continued RV monitoring will help constrain the periods and masses.

\begin{figure*}
    \centering
    \includegraphics[width=\textwidth]{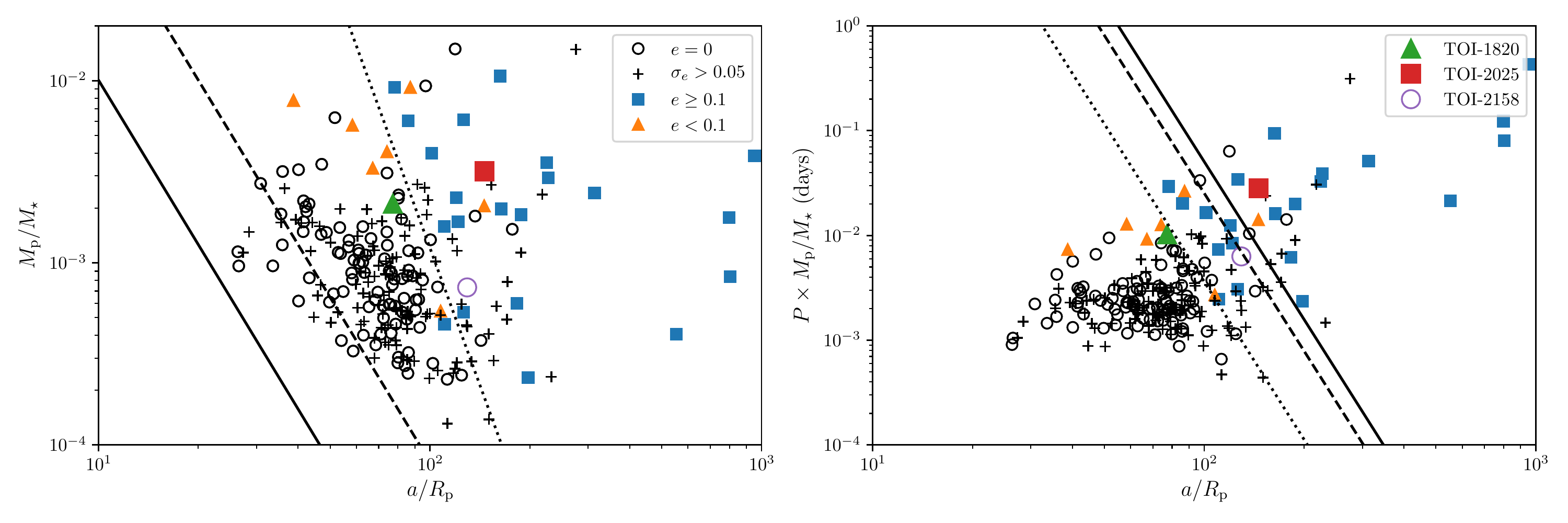}
    \caption{{\bf Tidal diagrams.} Tidal diagrams for transiting giant planets from \citet{Bonomo2017}. Open circles denote planets on circular orbits with $\sigma_e < 0.05$. Markers shown with pluses are planets with undertermined eccentrities, i.e. $\sigma_e > 0.05$. Most of these are consistent with $e=0$. Triangles represent planets with significant, but small eccentricities $e<0.1$, and squares are eccentric systems $\geq 0.1$. Adhering to this notation we have shown the planets in our sample with the corresponding marker. However, we have colour coded them for clarity. Created from the catalogue \citet{Bonomo2017Cat}. {\it Left: Tidal diagram.}  The solid and dashed lines show the position of a planet with a separation of $a=a_\mathrm{R}$ and $a=2a_\mathrm{R}$, respectively ($a_\mathrm{R}$ being the Roche limit), and radius $R_\mathrm{p}=1.2$~R$_\mathrm{J}$. The dotted line is a circularisation isochrone for a planet with $P=3$~d, $Q^{\prime}_\mathrm{p}=10^6$, and $e=0$. It should be noted that Eq. (1) in \citet{Bonomo2017} used to create the dotted line should have $\pi$ in the numerator instead of the denominator (A.~S.~Bonomo private communication)}. {\it Right: Modified tidal diagram. The dotted, dashed, and solid lines denote the 1, 7, and 14~Gyr circularisation timescales, respectively, assuming $e=0$ and $Q^{\prime}_\mathrm{p}=10^6$. }
    \label{fig:tidal}
\end{figure*}

In \fref{fig:tidal} we show the tidal diagram (left) and modified tidal diagram (right) from \citet{Bonomo2017} with our measurements for TOI-1820~b, TOI-2025~b, and TOI-2158~b. We find that the orbital eccentricity of TOI-2158~b is consistent with $e=0$. This planet joins the small group of planets in \citet{Bonomo2017} with circular orbits and relatively large values for $a/a_{\rm R}$, $a_{\rm R}$ being the Roche limit. This would allude to disc migration. However, given the age of $8 \pm 1$~Gyr for TOI-2158, the orbit of the planet might have had sufficient time to circularise, should the migration have taken place through high-eccentricity migration. For TOI-1820~b we find a modest eccentricity of \ecceighteen (about three times that of Earth). In \fref{fig:tidal} the planets with modest eccentricities are found at various relative masses and various relative distances. From the modified tidal diagram, it appears that TOI-1820~b should have a circularisation timescale of around 1-2~Gyr, and with the age of $11 \pm 2$~Gyr for TOI-1820, this leaves plenty of time for the system to dampen the eccentricity in the case of high-eccentricity migration. However, this modest eccentricity is not irreconcilable with disc migration \citep{Dawson2018}. In contrast, TOI-2025~b belongs to the subgroup of systems with significant eccentricity. The planet TOI-2025~b is too massive for the star to effectively raise tides on the planet in order to circularise the orbit, meaning that the circularisation timescale is too long for the orbit to have been circularised \citep{Dawson2018}. The modified tidal diagram suggests that the circularisation timescale could be some $10$~Gyr, which is much longer than the age of $1.7 \pm 0.2$~Gyr for this system.

On the same token, the planet seems to be massive enough for it to effectively raise tides on the star, while the star is sufficiently cool for tidal dissipation to be efficient \citep{Winn2010,Albrecht12b}. The projected obliquity we find for TOI-2025 is in line with other massive planets on eccentric, aligned orbits, such as HD~147506b \citep{Winn2007}, HD~17156~b \citep{Narita2009}, and HAT-P-34~b \citep{Albrecht12b}. Contrary to these findings, \citet{RiceWangLaughlin2022} has found that cool stars ($T_\mathrm{eff}<$6100~K) harbouring eccentric planets tend to have higher obliquities. Although, due to the sample size it is still unclear whether misalignment is associated with orbital eccentricity. Given the orbital, stellar, and planetary parameters, the low projected obliquity in TOI-2025 might be the result of tidal alignment \citep{Albrecht2022}. If so it would be interesting to further reduce the uncertainty of the obliquity measurement to test if the system is aligned to within $1^\circ$ as recently observed in some systems \citep{Albrecht2022}. This would suggest tidal alignment, as primordial alignment would presumably lead to a certain spread, as it has apparently done in the Solar System. TOI-1820 and TOI-2158 would, for similar reasons, be excellent RM targets as well. In addition, their higher impact parameters might lead to an even higher accuracy.

\section{Acknowledgements}
The authors would like to thank the referee, Louise~D.~Nielsen, for an insightful and helpful review of this work.
The authors would also like to thank the staff at the Nordic Optical Telescope for their help and expertise.
This paper includes data taken at the Nordic Optical Telescope under the programs IDs 59-210, 59-503, 61-510, 61-804, 62-506, and 63-505.
This study is based on observations made with the Nordic Optical Telescope, owned in collaboration by the University of Turku and Aarhus University, and operated jointly by Aarhus University, the University of Turku and the University of Oslo, representing Denmark, Finland, and Norway, the University of Iceland and Stockholm University at the Observatorio del Roque de los Muchachos, La Palma, Spain, of the Instituto de Astrofisica de Canarias.
This paper includes data taken at The McDonald Observatory of The University of Texas at Austin. This is University of Texas Center for Planetary Systems Habitability contribution \#0053.
We acknowledge the use of public TESS data from pipelines at the TESS Science Office and at the TESS Science Processing Operations Center. Resources supporting this work were provided by the NASA High-End Computing (HEC) programme through the NASA Advanced Supercomputing (NAS) Division at Ames Research Center for the production of the SPOC data products.
Funding for the Stellar Astrophysics Centre is provided by The Danish National Research Foundation (Grant agreement no.: DNRF106).
A.A.B., B.S.S., and I.A.S. acknowledge the support of Ministry of Science and Higher Education of the Russian Federation under the grant 075-15-2020-780(N13.1902.21.0039).
The numerical results presented in this work were obtained at the Centre for Scientific Computing, Aarhus \url{http://phys.au.dk/forskning/cscaa/}.
This work makes use of observations from the LCOGT network. Part of the LCOGT telescope time was granted by NOIRLab through the Mid-Scale Innovations Program (MSIP). MSIP is funded by NSF.
P. R. and L. M. acknowledge support from National Science Foundation grant No. 1952545.
This research made use of Astropy,\footnote{http://www.astropy.org} a community-developed core Python package for Astronomy \citep{art:astropy2013,art:astropy2018}. 
This research made use of matplotlib \citep{misc:hunter2007}.
This research made use of TESScut \citep{art:brasseur2019}.
This research made use of astroplan \citep{misc:morris2018}.
This research made use of SciPy \citep{misc:scipy2020}.
This research made use of corner \citep{corner}.
\bibliographystyle{aa} 
\bibliography{Bibliography} 

\begin{appendix}
\section{Tables for RVs, priors, and limb-darkening coefficients}

\begin{table*}[h!]
\begin{threeparttable}
 \centering
 \caption{{\bf Ground-based photometry.} 
 }
 \label{table:transitfollowup}
    \begin{tabularx}{\linewidth}{@{\extracolsep{\fill}}lclcccl}
    \toprule
    \toprule
Observatory  & Aperture (m) & Location                 & UTC Date      & Filter                  & Coverage  &   Planet\\
    \midrule
LCOGT\tnote{1}-Hal     & 0.4  & Haleakala, HI, USA        & 2020-05-06   &  Sloan $i'$              &  Ingress       &  TOI-1820 b\\
LCOGT-SSO              & 1.0  & Siding Spring, Australia  & 2020-05-11   &  $z$-short\tnote{2}       &  Egress        &  TOI-1820 b\\
LCOGT-McD              & 1.0  & McDonald Observatory, TX, USA & 2021-02-12 & $B$                       &  Full          &  TOI-1820 b\\
LCOGT-McD              & 1.0  & McDonald Observatory, TX, USA & 2021-02-12 & $z$-short               &  Full          &  TOI-1820 b\\
Kotizarovci            & 0.3  & Viskovo, Croatia              & 2020-06-26 & Baader $R$\tnote{3}       &  Full          & TOI-2025 b\\
LCOGT-TFN              & 0.4  & Tenerife, Canary Islands   & 2020-06-26    & Sloan $g'$              & Egress         & TOI-2025 b\\
LCOGT-TFN              & 0.4  & Tenerife, Canary Islands   & 2020-06-26    & Sloan $i'$              & Egress         & TOI-2025 b\\
FLWO\tnote{4}-KeplerCam  & 1.2  & Amado, Arizona, USA      & 2021-05-12    & $B$                       & Egress         & TOI-2025 b\\
FLWO-KeplerCam         & 1.2  & Amado, Arizona, USA         & 2021-05-12    & Sloan $i'$             & Egress         & TOI-2025 b\\
CRCAO-KU               & 0.6  & Kutztown, PA, USA          & 2021-05-21    & $R$                       & Full           & TOI-2025 b\\
Conti Private Obs.     & 0.3  & Annapolis, MD, USA         & 2021-12-19     & $V$                      & Full           & TOI-2025 b\\
LCOGT-McD              & 0.4  & McDonald Observatory, TX, USA & 2020-08-06 & Sloan $i'$              & Full           & TOI-2158 b\\
LCOGT-SAAO             & 1.0  & Cape Town, South Africa   & 2021-06-05   &  $B$                        & Full           & TOI-2158 b \\
LCOGT-SAAO             & 1.0  & Cape Town, South Africa   & 2021-06-05   &  $z$-short                & Full           & TOI-2158 b \\
 
    \bottomrule
    \bottomrule
    \end{tabularx}

\begin{tablenotes}
    \item Information on our ground-based photometric observations.
       \item [1] Las Cumbres Observatory Global Telescope \citep{Brown:2013}.
       \item [2] Pan-STARRS $z$-short.
       \item [3] Baader $R$ longpass 610~nm.
       \item [4] Fred L. Whipple Observatory.
     \end{tablenotes}
  \end{threeparttable}
\end{table*}

\begin{table}
    \centering
    \caption{{\bf Radial velocities for TOI-1820.} }
    \begin{threeparttable}
    \begin{tabular}{c c c c}
        \toprule
        Epoch (BJD$_\mathrm{TDB}$) & RV (m~s$^{-1}$) & $\sigma$ (m~s$^{-1}$) & Instrument \\
        \midrule
        2459269.780732 & 13841.9 & 12.6 & Tull \\ 
        2459270.880079 & 14196.4 & 14.9 & Tull \\ 
        2459275.91069 & 14205.8 & 15.7 & Tull \\ 
        2459276.9072 & 14101.6 & 14.6 & Tull \\ 
        2459277.930989 & 13785.6 & 10.4 & Tull \\ 
        2459280.881956 & 14225.0 & 8.8 & Tull \\ 
        2459281.790658 & 14083.9 & 16.3 & Tull \\ 
        2459293.694114 & 13727.5 & 11.3 & Tull \\ 
        2459294.83763 & 14080.0 & 13.3 & Tull \\ 
        2459301.873852 & 13885.5 & 34.0 & Tull \\ 
        2459302.781944 & 13685.8 & 13.0 & Tull \\ 
        2459308.757963 & 13845.4 & 9.8 & Tull \\ 
        2458991.51119436 & -0.0 & 18.9 & FIES \\ 
        2459204.77744922 & 128.2 & 16.8 & FIES \\ 
        2459206.67557103 & 130.2 & 27.7 & FIES \\ 
        2459215.77997042 & -20.9 & 21.5 & FIES \\ 
        2459229.68962246 & -63.7 & 35.2 & FIES \\ 
        2459233.68191458 & 231.9 & 17.5 & FIES \\ 
        2459235.67276872 & 86.3 & 19.8 & FIES \\ 
        2459239.60737547 & -59.5 & 11.8 & FIES \\ 
        2459243.67146977 & 119.2 & 12.1 & FIES \\ 
        2459254.55706331 & 12.6 & 16.4 & FIES \\ 
        2459276.67782068 & 453.3 & 15.7 & FIES \\ 
        2459277.72833782 & 107.7 & 15.7 & FIES \\ 
        2459283.69993517 & -28.2 & 15.6 & FIES \\ 
        2459284.68431354 & 198.0 & 31.6 & FIES \\ 
        2459285.67225885 & 495.6 & 15.3 & FIES \\ 
        2459290.64256983 & 507.0 & 15.5 & FIES \\ 
        2459291.542424 & 357.2 & 14.0 & FIES \\ 
        2459292.50793658 & 48.9 & 14.6 & FIES \\ 
        \bottomrule
    \end{tabular}
    \label{tab:rv_toi1820}
   \begin{tablenotes} 
    \item The epoch, RVs, and errors from our RV monitoring of TOI-1820.
     \end{tablenotes}
  \end{threeparttable}
\end{table}

\begin{table}
    \centering
    \caption{{\bf Radial velocities for TOI-2025.} }
    \begin{threeparttable}
    \begin{tabular}{c c c c}
        \toprule
        Epoch (BJD$_\mathrm{TDB}$) & RV (m~s$^{-1}$) & $\sigma$ (m~s$^{-1}$) & Instrument \\
        \midrule
2459124.41436227 & 2.5 & 33.9 & FIES \\ 
2459130.45161579 & -284.4 & 42.3 & FIES \\ 
2459131.34610963 & -183.9 & 38.9 & FIES \\ 
2459135.35261256 & -840.4 & 174.3 & FIES \\ 
2459138.35644793 & -379.9 & 39.6 & FIES \\ 
2459146.37902542 & -444.9 & 47.7 & FIES \\ 
2459147.40383093 & -333.3 & 32.7 & FIES \\ 
2459148.33080998 & -216.8 & 36.1 & FIES \\ 
2459149.32822725 & -11.5 & 24.1 & FIES \\ 
2459153.43955164 & -605.2 & 38.3 & FIES \\ 
2459156.34166702 & -245.9 & 136.3 & FIES \\ 
2459157.36143997 & -307.9 & 39.7 & FIES \\ 
2459162.3066763 & -682.9 & 29.3 & FIES \\ 
2459170.33208575 & -721.1 & 75.6 & FIES \\ 
2459201.80136496 & -85.7 & 36.0 & FIES \\ 
2459243.76155035 & -351.8 & 32.2 & FIES \\ 
2459255.74010029 & 25.2 & 31.6 & FIES \\ 
2459257.78775181 & -33.9 & 34.1 & FIES \\ 
2459367.59501228 & -350.7 & 20.3 & FIES \\ 
2459368.54449905 & -121.1 & 25.9 & FIES \\ 
2459420.48398551 & 2.2 & 17.3 & FIES+ \\ 
2459421.4708797 & 104.9 & 25.5 & FIES+ \\ 
2459424.55064799 & 488.6 & 17.1 & FIES+ \\ 
2459425.54286536 & 591.6 & 21.2 & FIES+ \\ 
2459442.62870736 & 541.1 & 27.8 & FIES+ \\ 
2459447.57636978 & 82.3 & 19.3 & FIES+ \\ 
2459465.4399404 & 43.1 & 14.5 & FIES+ \\ 
2459469.42183232 & 597.7 & 23.0 & FIES+ \\ 
2459473.38320428 & -59.0 & 18.8 & FIES+ \\ 
2459477.44240997 & 488.6 & 17.6 & FIES+ \\ 
2459435.4210958 & 296.7 & 17.1 & FIES+ \\ 
2459435.43653977 & 288.1 & 22.1 & FIES+ \\ 
2459435.45173373 & 257.1 & 17.8 & FIES+ \\ 
2459435.46695905 & 215.0 & 18.6 & FIES+ \\ 
2459435.48219576 & 231.8 & 28.6 & FIES+ \\ 
2459435.49740642 & 258.4 & 20.6 & FIES+ \\ 
2459435.51261722 & 258.6 & 15.3 & FIES+ \\ 
2459435.52786928 & 205.3 & 28.5 & FIES+ \\ 
2459435.54310052 & 166.6 & 18.7 & FIES+ \\ 
2459435.5582903 & 182.1 & 23.7 & FIES+ \\ 
2459435.57347547 & 140.3 & 20.8 & FIES+ \\ 
2459435.58863805 & 87.2 & 23.0 & FIES+ \\ 
2459435.60382369 & 79.3 & 24.0 & FIES+ \\ 
2459435.61903594 & 182.0 & 21.4 & FIES+ \\ 
2459435.63457285 & 147.6 & 17.9 & FIES+ \\ 
2459435.65079739 & 85.8 & 26.2 & FIES+ \\ 
2459614.77148284 & -118.2 & 50.0 & FIES+ \\ 
2459622.72491642 & -155.3 & 17.7 & FIES+ \\ 
2459623.7179694 & -171.9 & 24.4 & FIES+ \\ 
2459624.73548358 & -5.4 & 17.8 & FIES+ \\ 
2459647.6942654 & 540.3 & 59.7 & FIES+ \\ 
2459663.68611219 & 419.4 & 27.8 & FIES+ \\ 
2459664.68704606 & 583.9 & 20.3 & FIES+ \\ 
2459728.62936384 & -29.3 & 9.6 & FIES+ \\ 
2459732.6047247 & 197.9 & 14.3 & FIES+ \\ 
        \bottomrule
    \end{tabular}
    \label{tab:rv_toi2025}
   \begin{tablenotes} 
    \item The epoch, RVs, and errors from our RV monitoring of TOI-2025.
     \end{tablenotes}
  \end{threeparttable}
\end{table}

\begin{table}[h]
    \centering
    \caption{{\bf FIES radial velocities for TOI-2158.} }
    \begin{threeparttable}
    \begin{tabular}{c c c c}
        \toprule
        Epoch (BJD$_\mathrm{TDB}$) & RV (m~s$^{-1}$) & $\sigma$ (m~s$^{-1}$) & Instrument \\
        \midrule
2459332.67780456 & 0.1 & 10.5 & FIES \\ 
2459333.70360452 & -12.1 & 23.8 & FIES \\ 
2459339.681805 & -37.9 & 8.7 & FIES \\ 
2459351.64286508 & 123.2 & 7.8 & FIES \\ 
2459355.66056274 & 24.0 & 9.2 & FIES \\ 
2459364.63370853 & -15.1 & 10.9 & FIES \\ 
2459365.60471439 & 4.7 & 7.2 & FIES \\ 
2459367.51570125 & 93.2 & 8.1 & FIES \\ 
2459367.64102824 & 71.2 & 11.0 & FIES \\ 
2459368.63519994 & 140.6 & 8.4 & FIES \\ 
2459369.68467584 & 111.0 & 18.1 & FIES \\ 
2459371.62623245 & 72.3 & 5.9 & FIES \\ 
2459372.49619057 & 19.1 & 14.2 & FIES \\ 
2459376.59299817 & 96.0 & 11.3 & FIES \\ 
2459380.64168281 & 31.2 & 15.6 & FIES \\ 
2459382.55110354 & -7.0 & 6.8 & FIES \\ 
2459425.45350272 & -0.0 & 5.1 & FIES+ \\ 
2459428.61860581 & 142.7 & 13.8 & FIES+ \\ 
2459434.45668389 & -15.6 & 4.6 & FIES+ \\ 
2459436.5266032 & 77.5 & 4.6 & FIES+ \\ 
2459438.44453163 & 127.4 & 5.7 & FIES+ \\ 
2459442.44074223 & -24.5 & 5.3 & FIES+ \\ 
2459447.49151578 & 141.4 & 4.7 & FIES+ \\ 
2459449.40026218 & 42.0 & 4.7 & FIES+ \\ 
2459451.55658931 & -8.9 & 5.2 & FIES+ \\ 
2459453.507784 & 76.1 & 3.9 & FIES+ \\ 
2459464.43808474 & 151.5 & 4.2 & FIES+ \\ 
2459465.48015278 & 91.3 & 5.2 & FIES+ \\ 
2459467.45855826 & 7.2 & 4.3 & FIES+ \\ 
2459470.45543094 & 47.9 & 13.4 & FIES+ \\ 
2459477.40400508 & -5.7 & 4.8 & FIES+ \\ 
2459622.7602177 & -50.6 & 9.1 & FIES+ \\ 
2459623.75257858 & -36.6 & 5.7 & FIES+ \\ 
2459627.76305009 & 115.6 & 10.1 & FIES+ \\ 
2459633.74209536 & 48.6 & 14.6 & FIES+ \\ 
2459663.7250026 & 49.6 & 6.4 & FIES+ \\ 
2459664.72649871 & -27.3 & 8.8 & FIES+ \\ 
2459733.59894098 & -31.8 & 4.9 & FIES+ \\

        \bottomrule
    \end{tabular}
    \label{tab:rv_toi2158a}
   \begin{tablenotes} 
    \item The epoch, RVs, and errors from our FIES RV monitoring of TOI-2158.
     \end{tablenotes}
  \end{threeparttable}
\end{table}

\begin{table}[h]
    \centering
    \caption{{\bf Tull radial velocities for TOI-2158.} }
    \begin{threeparttable}
    \begin{tabular}{c c c c}
        \toprule
        Epoch (BJD$_\mathrm{TDB}$) & RV (m~s$^{-1}$) & $\sigma$ (m~s$^{-1}$) & Instrument \\
        \midrule

2459302.925697 & -64804.4 & 23.8 & Tull \\ 
2459308.916426 & -64749.6 & 20.7 & Tull \\ 
2459309.847096 & -64745.0 & 23.6 & Tull \\ 
2459339.927982 & -64807.7 & 22.9 & Tull \\ 
2459340.803182 & -64804.8 & 21.5 & Tull \\ 
2459355.902204 & -64841.0 & 22.7 & Tull \\ 
2459372.717338 & -64779.2 & 22.0 & Tull \\ 
2459384.739579 & -64731.2 & 21.2 & Tull \\ 
2459385.864151 & -64672.0 & 23.6 & Tull \\ 
2459411.718672 & -64646.1 & 20.9 & Tull \\ 
2459412.852352 & -64662.1 & 23.1 & Tull \\ 
2459413.7245 & -64636.3 & 23.7 & Tull \\ 
2459454.684757 & -64668.7 & 22.9 & Tull \\ 
2459455.705819 & -64582.0 & 22.9 & Tull \\ 
2459456.718945 & -64660.5 & 24.8 & Tull \\ 
2459471.635406 & -64623.3 & 25.2 & Tull \\ 
2459472.60132 & -64609.5 & 23.6 & Tull \\ 
2459514.575615 & -64662.9 & 22.8 & Tull \\ 
2459515.606634 & -64660.8 & 23.8 & Tull \\ 
2459516.580362 & -64647.4 & 23.2 & Tull \\ 
2459528.573076 & -64735.0 & 22.6 & Tull \\ 
2459529.557203 & -64775.8 & 22.2 & Tull \\ 
2459541.556578 & -64646.5 & 25.7 & Tull \\ 
2459610.026931 & -64665.1 & 29.2 & Tull \\ 
2459611.019771 & -64694.2 & 26.6 & Tull \\ 
2459622.022527 & -64851.4 & 24.6 & Tull \\ 
2459634.984303 & -64685.8 & 27.5 & Tull \\ 
2459692.973827 & -64845.7 & 24.9 & Tull \\ 
2459693.918368 & -64826.4 & 22.2 & Tull \\ 
2459705.915839 & -64728.1 & 19.6 & Tull \\ 
2459706.844475 & -64782.3 & 20.9 & Tull \\ 
2459707.907303 & -64839.9 & 23.4 & Tull \\ 
2459707.922283 & -64835.6 & 23.7 & Tull \\

        \bottomrule
    \end{tabular}
    \label{tab:rv_toi2158b}
   \begin{tablenotes} 
    \item The epoch, RVs, and errors from our Tull RV monitoring of TOI-2158.
     \end{tablenotes}
  \end{threeparttable}
\end{table}

\begin{table}[h]
    \centering
    \caption{ {\bf Priors used in our MCMC.} }
    \begin{threeparttable}
    \begin{tabular}{c c c c}
    \toprule
    Parameter & TOI-1820 & TOI-2025 & TOI-2158 \\
    \midrule
    
$P$& $\mathcal{U}$ & $\mathcal{U}$ & $\mathcal{U}$  \\ 
 $T_0$& $\mathcal{U}$ & $\mathcal{U}$ & $\mathcal{U}$  \\ 
 $R_\mathrm{p}/R_\star$& $\mathcal{U}$ & $\mathcal{U}$ & $\mathcal{U}$  \\ 
 $a/R_\star$& $\mathcal{U}$ & $\mathcal{U}$ & $\mathcal{U}$  \\ 
 $K$& $\mathcal{U}$ & $\mathcal{U}$ & $\mathcal{U}$  \\ 
 $\cos i$& $\mathcal{U}$ & $\mathcal{U}$ & $\mathcal{U}$  \\ 
 $\sqrt{e} \cos \omega$& $\mathcal{U}$ & $\mathcal{U}$ & $\mathcal{U}$  \\ 
 $\sqrt{e} \sin \omega$& $\mathcal{U}$ & $\mathcal{U}$ & $\mathcal{U}$  \\ 
 $\lambda$ & - & $\mathcal{U}$  & -  \\ 
 $\gamma_1$& $\mathcal{U}$ & $\mathcal{U}$ & $\mathcal{U}$  \\ 
 $\gamma_2$& - & $\mathcal{U}$ & $\mathcal{U}$  \\ 
 $\gamma_3$ & $\mathcal{U}$  & - & $\mathcal{U}$  \\ 
 $\sigma_1$& $\mathcal{U}$ & $\mathcal{U}$ & $\mathcal{U}$  \\ 
 $\sigma_2$& - & $\mathcal{U}$ & $\mathcal{U}$  \\ 
 $\sigma_3$ & $\mathcal{U}$  & - & $\mathcal{U}$  \\ 
 $\log A_1$ & - & $\mathcal{U}$ & $\mathcal{U}$  \\ 
 $\log \tau_1$ & - & $\mathcal{U}$ & $\mathcal{U}$  \\ 
 $\log A_2$& $\mathcal{U}$ & $\mathcal{U}$ & $\mathcal{U}$  \\ 
 $\log \tau_2$& $\mathcal{U}$ & $\mathcal{U}$ & $\mathcal{U}$  \\ 
 $\ddot{\gamma}$ & - & $\mathcal{U}$ & $\mathcal{U}$  \\ 
 $\dot{\gamma}$ & - & $\mathcal{U}$ & $\mathcal{U}$  \\ 
 $\delta \mathrm{M}_I$ & $\mathcal{N}(4.0,0.5)$  & -  & -  \\ 
 $\delta \mathrm{M}_B$ & $\mathcal{N}(4.5,1.0)$  & -  & -  \\ 
 $v \sin i_\star$ & -  & $\mathcal{N}(6.0,0.3)$  & -  \\ 
 $\zeta$ & -  & $\mathcal{N}(4,1)$  & -  \\ 
 $\xi$ & -  & $\mathcal{N}(1,1)$  & -  \\ 
 
 \bottomrule
    \end{tabular}

    \label{tab:priors}
   \begin{tablenotes} 
    \item $\mathcal{U}$ denotes a uniform prior and $\mathcal{N}(\mu,\sigma)$ is a Gaussian prior with mean $\mu$ and standard deviation $\sigma$. Description of the parameters can be found in \tref{tab:mcmc}.
     \end{tablenotes}
  \end{threeparttable}

\end{table}

\begin{table}
    \centering
    \caption{{\bf Limb-darkening coefficients for TOI-1820.} }
    \begin{threeparttable}
    \begin{tabular}{c c c c c}
    \toprule
         &  \multicolumn{4}{c}{TOI-1820} \\
         & \multicolumn{2}{c}{Initial} & \multicolumn{2}{c}{Results} \\
         & $q_1$ & $q_2$ & $q_1$ & $q_2$ \\
    \midrule
{\it TESS} & 0.2986 & 0.2806 & $0.26\pm0.03$ & $0.25\pm0.03$ \\ 
 LCO HAL $i^{\prime}$ & 0.3815 & 0.1936 & $0.38_{-0.05}^{+0.04}$ & $0.19_{-0.05}^{+0.04}$ \\ 
 LCO McD $B$ & 0.7104 & 0.1172 & $0.81_{-0.03}^{+0.04}$ & $0.21_{-0.03}^{+0.04}$ \\ 
 LCO McD $z$-short & 0.3152 & 0.1968 & $0.33\pm0.04$ & $0.21\pm0.04$ \\ 
 LCO SSO $z$-short & 0.3152 & 0.1968 & $0.32\pm0.05$ & $0.21\pm0.05$ \\ 
 
    \bottomrule
    \end{tabular}
   \begin{tablenotes} 
    \item The initial and resulting values for the linear, $q_1$, and quadratic, $q_2$, limb-darkening coefficients from the different photometric systems. In our MCMC we step in the sum, $q_1+q_2$, while applying a Gaussian prior with a width of 0.1 and keeping the difference, $q_1-q_2$, fixed.
     \end{tablenotes}
  \end{threeparttable}    
    \label{tab:ld_toi1820}
\end{table}

\begin{table}
    \centering
    \caption{{\bf Limb-darkening coefficients for TOI-2025.} }
    \begin{threeparttable}
    \begin{tabular}{c c c c c}
    \toprule
         &  \multicolumn{4}{c}{TOI-2025} \\
         & \multicolumn{2}{c}{Initial} & \multicolumn{2}{c}{Results} \\
        & $q_1$ & $q_2$ & $q_1$ & $q_2$ \\
    \midrule
{\it TESS} & 0.263 & 0.2978 & $0.22\pm0.02$ & $0.26\pm0.02$ \\ 
 {\it Kepler} $i^{\prime}$ & 0.3815 & 0.1936 & $0.38\pm0.05$ & $0.19\pm0.05$ \\ 
 {\it Kepler} $B$ & 0.7104 & 0.1172 & $0.70\pm0.04$ & $0.10\pm0.04$ \\ 
 LCO TFN $i^{\prime}$ & 0.3815 & 0.1936 & $0.37\pm0.05$ & $0.18\pm0.05$ \\ 
 LCO TFN $g^{\prime}$ & 0.6533 & 0.1337 & $0.66\pm0.05$ & $0.14\pm0.05$ \\ 
 Conti $V$ & 0.5501 & 0.177 & $0.55\pm0.05$ & $0.18\pm0.05$ \\ 
 CRCAO $R$ & 0.4566 & 0.1867 & $0.45_{-0.05}^{+0.04}$ & $0.18_{-0.05}^{+0.04}$ \\ 
 Kotizarovci & 0.263 & 0.2978 & $0.24\pm0.05$ & $0.28\pm0.05$ \\ 
 FIES+ & 0.5501 & 0.1777 & $0.55\pm0.05$ & $0.18\pm0.05$ \\ 
 
    \bottomrule
    \end{tabular}
   \begin{tablenotes} 
    \item Same as in \tref{tab:ld_toi1820}.
     \end{tablenotes}
  \end{threeparttable}        
    \label{tab:ld_toi2025}
\end{table}

\begin{table}[]
    \centering
    \caption{{\bf Limb-darkening coefficients for TOI-2158.}}
    \begin{threeparttable}
    \begin{tabular}{c c c c c}
    \toprule
         &  \multicolumn{4}{c}{TOI-2158} \\
         & \multicolumn{2}{c}{Initial} & \multicolumn{2}{c}{Results} \\
         & $q_1$ & $q_2$ & $q_1$ & $q_2$ \\
    \midrule
{\it TESS} & 0.3249 & 0.2771 & $0.30\pm0.03$ & $0.25\pm0.03$ \\ 
 LCO McD $i^{\prime}$ & 0.7871 & 0.0485 & $0.86\pm0.04$ & $0.12\pm0.04$ \\ 
 LCO SAAO $B$ & 0.3365 & 0.1865 & $0.31_{-0.04}^{+0.05}$ & $0.16_{-0.04}^{+0.05}$ \\ 
 LCO SAAO $z$-short & 0.3249 & 0.2771 & $0.33\pm0.05$ & $0.28\pm0.05$ \\ 
 
    \bottomrule
    \end{tabular}
   \begin{tablenotes} 
    \item Same as in \tref{tab:ld_toi1820}.
     \end{tablenotes}
  \end{threeparttable}    
    \label{tab:ld_toi2158}
\end{table}

\section{Additional figures}
\begin{figure*}
    \centering
    \includegraphics[width=\textwidth]{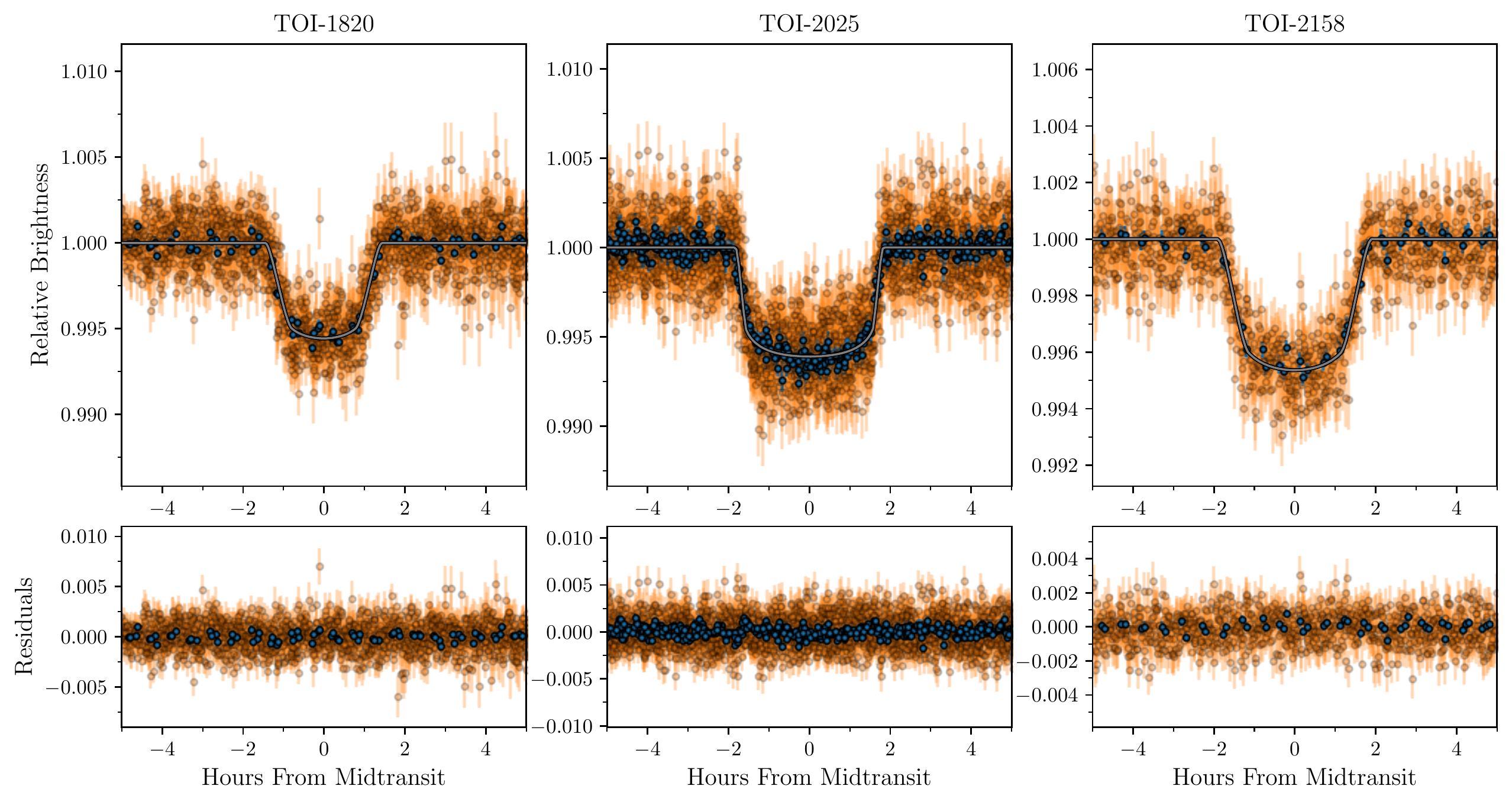}
    \caption{{\bf \tess curves}. A close-up of the \tess 2-min (orange) and 30-min (blue) cadence light curves for TOI-1820 (left), TOI-2025 (middle), and TOI-2158 (right). The residuals show no structure around the transits.}
    \label{fig:lc_tess2min}
\end{figure*}

\begin{figure}
    \centering
    \includegraphics[width=\columnwidth]{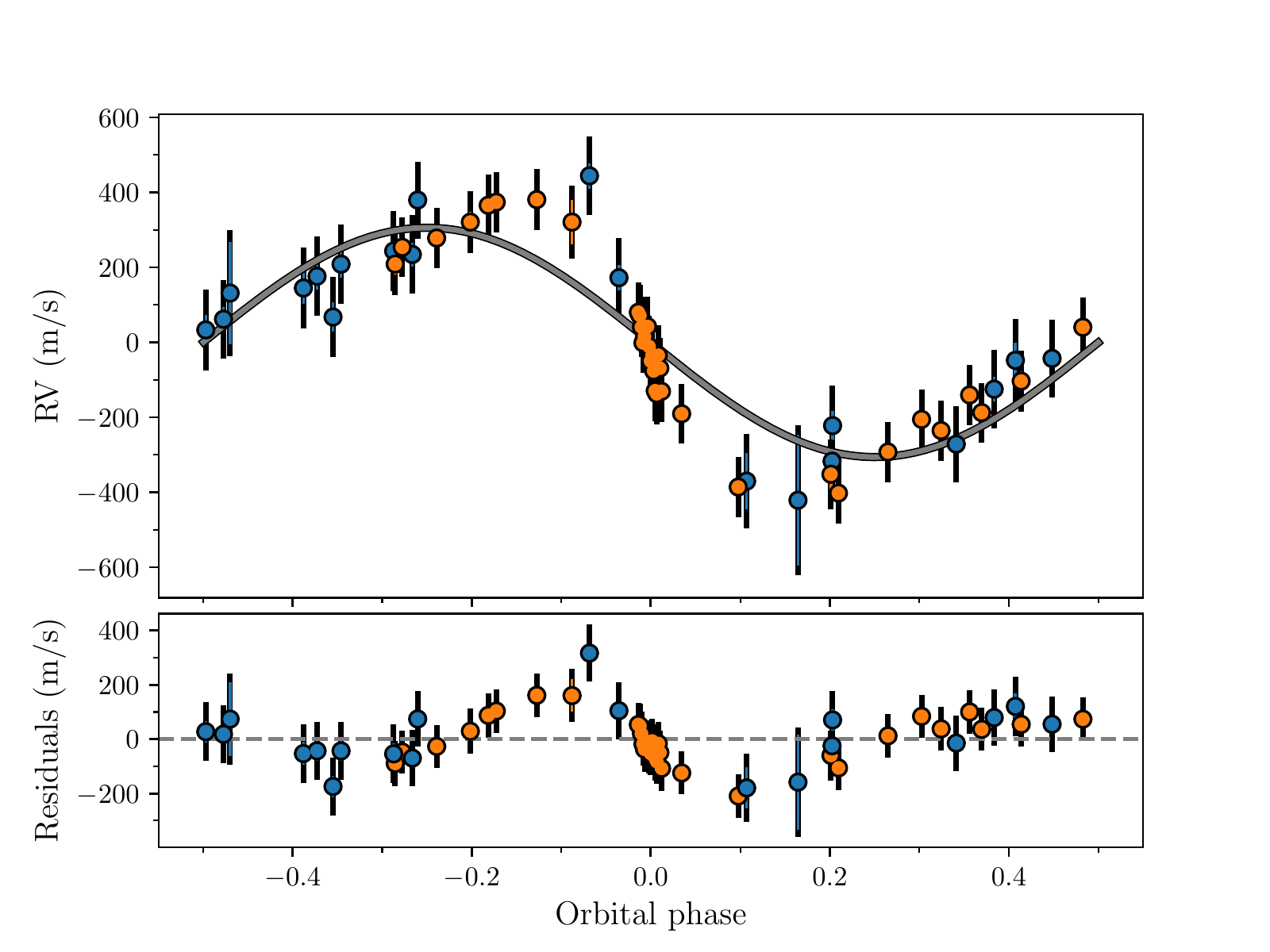}
    \caption{{\bf Circular solution for TOI-2025~b.} Symbols are the same as in the middle panel of \fref{fig:rv_all}, but here we have fixed $e=0$ during the MCMC. }
    \label{fig:rv_toi2025_no_ecc}
\end{figure}

\end{appendix}
\end{document}